\documentclass[reprint,amsmath,amssymb]{revtex4-2}
\usepackage[rgb]{xcolor}
\usepackage{tikz}
\usepackage{booktabs}

\usepackage{hyperref}
\newcommand{\be}{\begin{equation}}
\newcommand{\ee}{\end{equation}}
\newcommand{\beqn}{\begin{eqnarray}}
\newcommand{\eeqn}{\end{eqnarray}}
\usetikzlibrary{shapes.geometric}

\definecolor{myred}{rgb}{0.90, 0.36, 0.054}
\definecolor{myblue}{rgb}{0.365248, 0.427802, 0.758297}
\definecolor{myorange}{rgb}{0.945109, 0.593901, 0}
\definecolor{mygreen}{rgb}{0.285821, 0.56, 0.450773}
\definecolor{mypurple}{rgb}{0.645957, 0.253192, 0.685109}

\newcommand\drawcircle[1][black]{%
\begin{tikzpicture}
\fill[#1] (0,0) circle (3pt);
\end{tikzpicture}
} 

\newcommand\drawsquare[1][black]{%
\begin{tikzpicture}
\fill[#1] (0,0) rectangle (6pt,6pt);
\end{tikzpicture}
}

\begin{document}

\title{Cluster tomography in percolation}
\author{Helen S. Ansell}
\affiliation{Department of Physics and Astronomy, Northwestern University, Evanston, IL 60208}
\author{Samuel J. Frank}
\affiliation{Department of Physics and Astronomy, Northwestern University, Evanston, IL 60208}
\author{Istv\'an A. Kov\'acs}
\affiliation{Department of Physics and Astronomy, Northwestern University, Evanston, IL 60208}
\affiliation{Northwestern Institute on Complex Systems, Northwestern University, Evanston, IL 60208}
\affiliation{Department of Engineering Sciences and Applied Mathematics, Northwestern University, Evanston, IL 60208}

\date{\today}

\begin{abstract}
In cluster tomography, we propose measuring the number of clusters $N$ intersected by a line segment of length $\ell$ across a finite sample.
As expected, the leading order of $N(\ell)$ scales 
as $a\ell$, where $a$ depends on microscopic details of the system. However, at criticality, there is often an additional nonlinearity of the form $b\ln(\ell)$, originating from the endpoints of the line segment. By performing large scale Monte Carlo simulations of both 2$d$ and 3$d$ percolation, we find that $b$ is universal and depends only on the angles encountered at the endpoints of the line segment intersecting the sample. 
Our findings are further supported by analytic arguments in 2$d$, building on results in conformal field theory. 
Being broadly applicable, cluster tomography can be an efficient tool to detect phase transitions and to characterize the corresponding universality class in classical or quantum systems with a relevant cluster structure. 
\end{abstract}

\maketitle

Cluster formation is a prevalent feature of complex systems, including but not restricted to magnetic domains \cite{komogortsev2019}, motility-induced phase
separation \cite{tailleur2008, Buttinoni2013, Palacci2013}, bacteria swarming \cite{be'er2020}, cell migration \cite{swart2002, guo2002} and the collective motion of animal groups \cite{vicsek1995,cavagna2014, katz2011, gordon2014}. As phase transitions often lead to changes in the characteristics of the emerging clusters, phase transitions can --- in principle --- be detected and studied through cluster statistics. 
As a realization, let us consider a scenario 
where a probe is shot through a complex system consisting of clusters and ask the following question:
\emph{How many clusters are encountered by the probe during the measurement?}
Here we propose that such measures of ``cluster tomography'' yield simple and efficient methods for locating critical points in complex systems, both experimentally and computationally, while also providing deep universal information on the nature of the observed transitions.

\begin{figure}
\centering
\includegraphics[width=\linewidth]{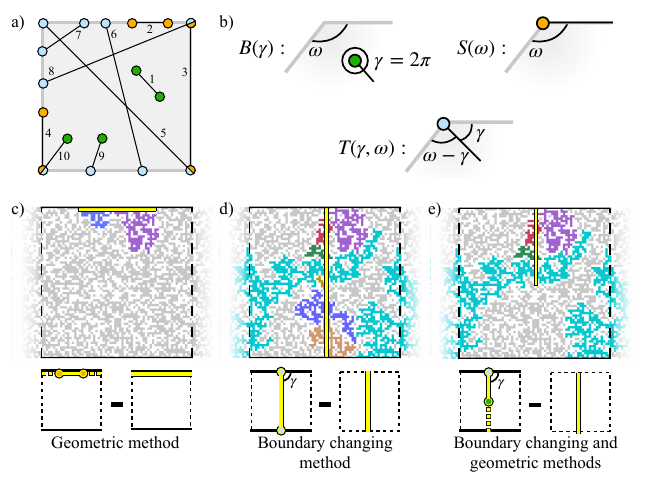}
\vskip -.5cm
\caption{
\label{fig:ill} 
(a) Summary of all line segment types 
in a 2$d$ square system with free boundaries. Endpoints of the lines are colored according to the classification in (b), which are bulk (green), surface (orange), and traversing (blue).
(c-e) Schematics of contributing clusters and numerical techniques used to access the critical corner contribution of different line segments for site percolation, illustrated for line types 2, 6 and 9.
Dashed black lines represent PBC, solid lines represent FBC, and solid yellow lines represent the measured line segment. 
Sites belonging to clusters that are counted by the line segment are colored while the remaining occupied sites are gray. 
}
\end{figure}

In this letter, we first address all ten configuration types of cluster tomography in a two-dimensional (2$d$) square system, as illustrated in Fig.~\ref{fig:ill}. 
We focus on Bernoulli percolation on a square lattice, although the main results are expected to apply more generally \cite{kovacs14}, as also illustrated for 3$d$ percolation.
Percolation is a fundamental model of critical phenomena, where sites or bonds of a lattice are independently occupied with probability $p$. Collective behavior emerges as we are interested in the statistics of clusters of connected sites \cite{stauffer}. In dimensions $d\geq2$ there is a scale-invariant critical point at $p=p_c$ at which the system is also believed to be conformally invariant \cite{smirnov,*tsai2013}. In the most studied 2$d$ case, there are many exact results available \cite{conf_inv, sle,*SmirnovWerner2001}, including correlation functions \cite{dotsenko_fateev}, crossing probabilities \cite{crossing} and critical exponents. 

Motivated by questions on quantum entanglement in disordered systems \cite{yu07, kovacs, kovacs14}, previous studies have considered the number of clusters (magnetic domains), $N_{\Gamma}$ intersecting a contour, $\Gamma$. Cluster tomography corresponds to the simplest case of skeletal entanglement \cite{skeletal} when $\Gamma$ is a line segment. For select line configurations of length $\ell$, it was found that at criticality 
\begin{equation}
N(\ell)=a\ell+b\ln(\ell)+O(1)\;, 
\label{N}
\end{equation}
in both 2$d$ and 3$d$ \cite{yu07, kovacs, kovacs3dperc, kovacs_largeQ}. 
The first term represents the expected ``area law'' scaling of the number of clusters with the length of the line, with $a$ being the non-universal linear cluster density.

The term $b\ln(\ell)$ is the ``corner contribution'' that emerges due to the geometric singularities at the endpoints of the line segment, akin to a notion of ``geometric susceptibility'' \cite{Witczak-Krempa2019}. This term is present only at criticality and 
is an elegant measure of the concavity of the underlying cluster geometry that
encapsulates universal information from correlations in the cluster shape at all orders.
For sufficiently long lines --- i.e., $\ell/L=O(1)$, with $L$ being the system size --- each endpoint is expected to contribute to the ``cluster count exponent'' $b$ independently.
For a given configuration, $b$ is expected to be universal, that is independent of the microscopic details of the model at criticality.
The value of $b$ depends on the endpoint configurations of the line segment, through the angle(s) appearing at each endpoint.  
If the endpoint touches the system boundary, the angle $\omega$ of the boundary at the endpoint also plays a role, see Fig.~\ref{fig:ill}b. 
For line types 1-5 this means that the value of $b$ is universal, while for lines 6-10 $b$ is a universal function of the characteristic angle $\gamma$.

The contribution of each endpoint depends on its topology, with three distinct cases in 2$d$: bulk $b$, surface $s$ and traversing $t$, as illustrated in Fig.~\ref{fig:ill}b. Bulk endpoints have $\gamma=2\pi$ and occur when the endpoint is not on the surface, while surface endpoints, for which the corner contribution depends on $\omega$, occur when the line segment is along the surface. Traversing endpoints occur when the endpoint is on the surface and the line segment crosses the bulk of the system. 
%
Supported by analytic arguments from conformal field theory (see Appendix A for details), 
the contributions of the endpoints consist of linear combinations of a bulk term $B$ and a surface term $S$, depending on the configuration. 
The bulk term 
	\begin{equation}
		 B(\gamma)=\frac{5\sqrt{3}}{96\pi}\Big(\frac{\gamma}{\pi}-\frac{\pi}{\gamma}\Big) 
		 \label{eq:B}
	\end{equation}
follows from the celebrated Cardy-Peschel formula \cite{cardypeschel}, which appears whenever the line segment passes through the bulk of the system.  
The surface term 
	\begin{equation}
		 S(\gamma)=\frac{\sqrt{3}}{8\pi}\frac{\pi}{\gamma}
		 \label{eq:S}
	\end{equation}	 
appears whenever there is a change of boundary condition at the endpoint \cite{CardyBC,stephan2013}. 
Such a change of boundary condition occurs when the line segment along which clusters are counted (fixed boundary) meets the surface (free boundary) at an angle, which occurs for type $s$ and $t$ endpoints.
Each surface endpoint contributes $S(\omega)$ to $b$; while bulk endpoints each contribute $B(2\pi)$. Note that usually a bulk corner contribution would be the sum of $B(\gamma)$ and that of the conjugate angle $B(2\pi-\gamma)$ \cite{kovacs, kovacs14}. 
However, for $\gamma=2\pi$, the contribution is just $B(2\pi)$, as the singular $B(0)$ is not sampled by the measurement \cite{kovacs, kovacs14}. 

Traversing endpoints are more complicated, as multiple angles contribute. 
The na\"ive expectation would account for the bulk contributions from the angle $\gamma$ and the complementary angle $\omega-\gamma$ between the measuring line and the surface, as well as surface contributions from these angles due to the  boundary condition change. However, the sum of the terms $B(\gamma)+S(\gamma)+B(\omega-\gamma)+S(\omega-\gamma)$ would correspond to a different configuration, where clusters are also counted along the free surface lines. Such counting along two lines at an angle $\omega$ would correspond to $B(\omega)$ on its own, as there is no boundary condition change (and no exterior angle). 
Therefore, the na\"ive formula is expected to be the sum of $B(\omega)$ and the contribution of the traversing endpoint, leading to each traversing endpoint contributing
\begin{equation}
{T}(\gamma,\omega)=B(\gamma)+S(\gamma)+B(\omega-\gamma)+S(\omega-\gamma) -B(\omega)\;.
\end{equation}
Note that since $B(\pi) = 0$, here the $-B(\omega)$ contribution to $T(\gamma,\omega)$ can only be detected if the endpoint is in the square corner ($\omega=\pi/2$), corresponding to a type $t_c$ endpoint.
Once again, $\gamma=0$ requires special care, as $ T(0,\omega)$ has unmeasured singular contributions, so $S(\omega)$ should be used directly in this case.
These arguments lead to our predicted expressions for the cluster count exponent $b_{ij}$ listed in Table~\ref{tab:lines}, where $i,j \in \{b, s, t\}$ denotes the endpoint type and an additional subscript $c$ is used on $s$ and $t$ to indicate if the endpoint is in the corner of the square system.
%

These predictions for 2$d$ percolation need to be tested numerically. Moreover, in other systems lacking detailed analytic results, numerical measurements are the only option to perform cluster tomography.
It is, however, not immediately obvious that the universal logarithmic corner contribution can be measured to high-precision. In practice, there are multiple difficulties to overcome, including the fact that the linear term $a\ell$ is much larger than the nonlinear term $b\ln(\ell)$, large statistical noise in the overall measurement $\sim O(\sqrt{\ell})$, and potentially strong finite-size corrections to the linear term. Strikingly, the area law term can be 
canceled out \emph{exactly} together with its statistical error and finite-size corrections, leading to precise measurements of the corner term even in relatively small systems. 

For line types 1 and 2, the corner contribution can be precisely measured using a \emph{geometric method} \cite{kovacs, kovacs3dperc}, as illustrated in Fig.~\ref{fig:ill}c for a line segment of length \(\ell = L/2\) on a free surface.
The difference between the total number of clusters on two contributing segments of length $\ell=L/2$ and the number of clusters on a full periodic line of length $L$ (without any endpoints) gives twice the corner contribution of the line segment. 
As the same sites are visited in both cases, the linear area law term cancels out and the only remaining contribution must come from the endpoints. 
The corner contribution is therefore simply half the number of shared clusters between the two line segments. 
This approach has previously been used to numerically calculate $b_{ss} $ in 2$d$, as well as $b_{bb}$ in 2$d$ and 3$d$ \cite{kovacs, kovacs3dperc}, using the same logic for a line segment in the bulk  (Fig.~\ref{fig:ill}a, line 1) with periodic boundary conditions (PBC).
These 2$d$ results are consistent with predictions from conformal field theory \cite{kovacs}, and the 2$d$ surface result has been proven rigorously on the triangular lattice \cite{berg}.

As the applicability of the geometric method alone is limited to these two cases, here we propose a \emph{boundary changing method} to measure the corner contribution of the remaining line types with high precision. 
This method is expected to be applicable for cluster-based systems with short-range interactions, and is illustrated in Fig.~\ref{fig:ill}d for a line segment with traversing endpoints on opposite sides of the system. 
By changing the free boundary conditions (FBC) into PBC, the same line now forms a closed loop with no endpoints. 
The area law term can be canceled by taking the difference between the number of clusters that intersect the line segment with FBC and PBC.  
The boundary changing method can be used alone, or in conjunction with the geometric method (as illustrated in Fig.~\ref{fig:ill}e), to numerically determine the corner contributions of line types 3-10, including the angle dependence of lines 6-10, as indicated in Table~\ref{tab:lines}.

\begin{table*}
    \centering
    \footnotesize
     \caption{Overview of results 
    for 2$d$ percolation on a square lattice. 
    Analytic and numerical values are given for the least acute angle studied $\gamma(n_{min})$ in cases where $b$ depends on $\gamma$.
    Parentheses by measured values give the measurement error in the last digit. 
   ``G'' and ``BC'' respectively refer to the geometric and boundary changing methods, * indicates cancellation of lines of type 6 and/or 9 is required to determine $b$ to high precision numerically.
    }
    \begin{tabular}{cccccccc}
    \toprule
    \toprule
    Line &Endpoints\footnote{($'$) for line 7 indicates endpoints are on adjacent edges} &Fig.  & Formula &$\gamma(n_{min})$ &$b$ expected & $b$ measured & Technique\\
    \midrule
    1 & $bb$ &\parbox[c]{2em}{\includegraphics[width=0.3in]{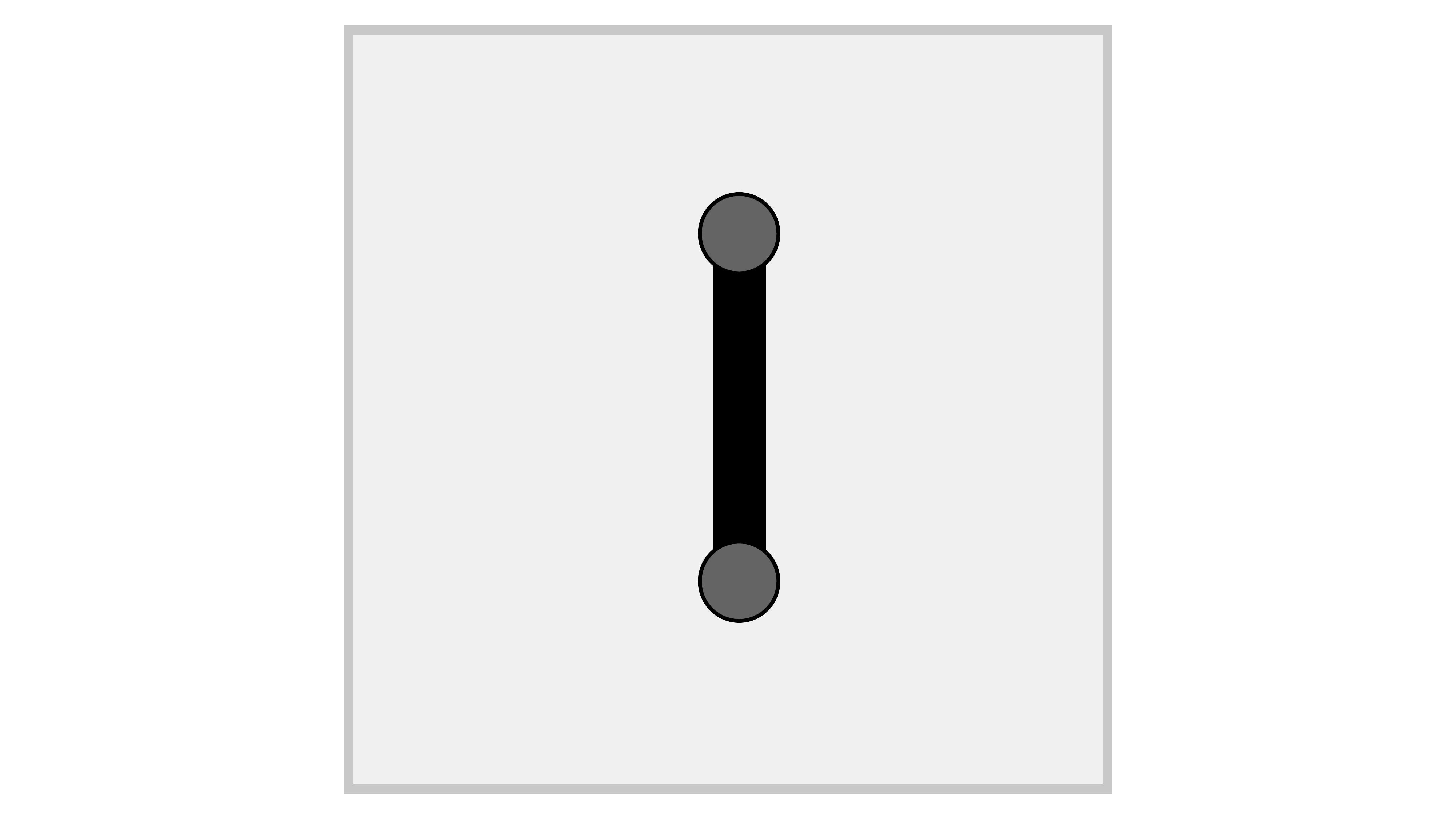}} & $2B(2\pi)$ & - &\(\frac{5\sqrt{3}}{32\pi}\approx 0.086\)\footnotemark[2]\footnotetext[2]{Reference \cite{kovacs}}& 0.086(1)\footnotemark[2] &G\\[3 pt]
         2 & $ss$ & \parbox[c]{2em}{\includegraphics[width=0.3in]{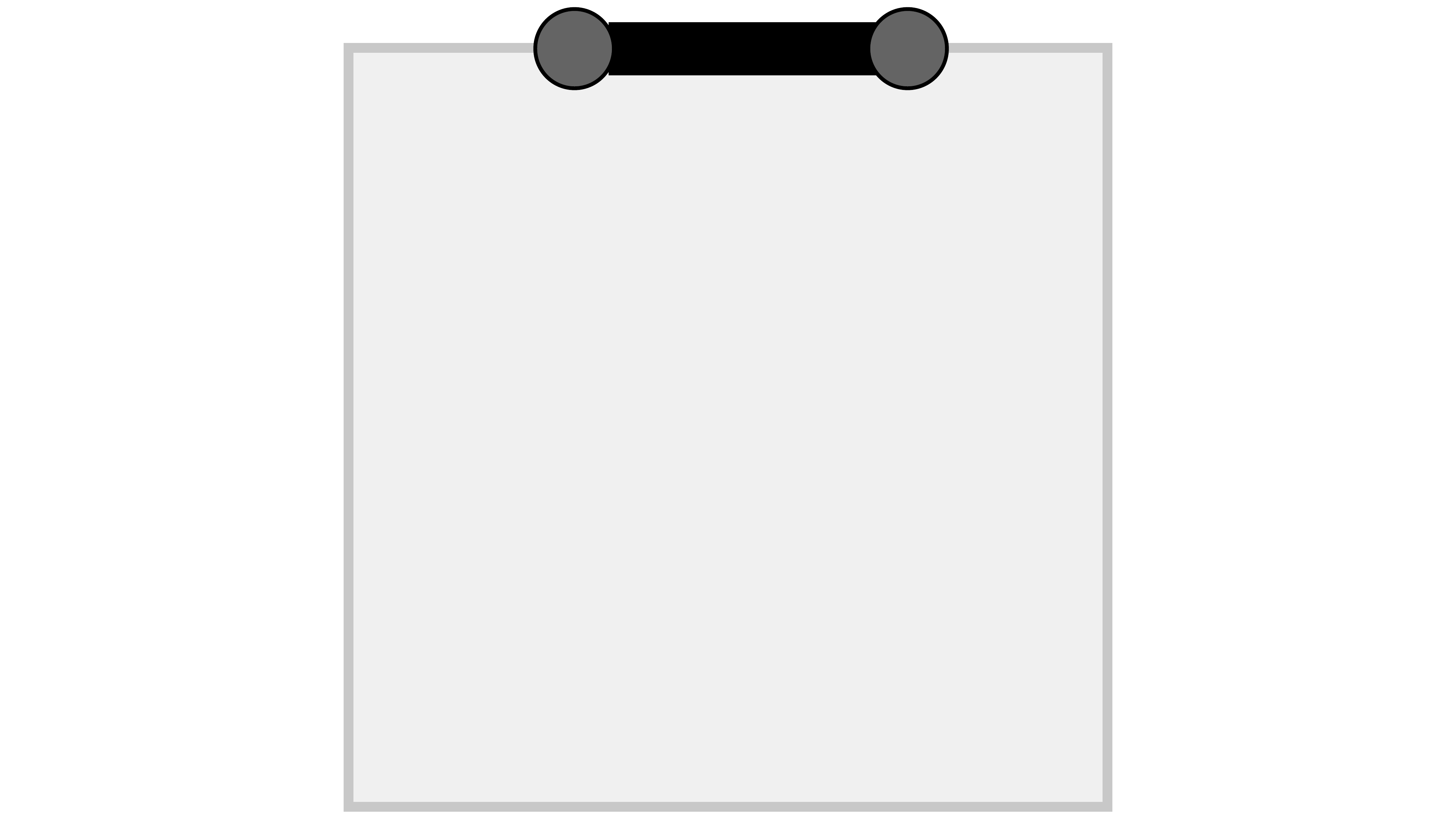}}  & $2S(\pi)$ & - & \(\frac{\sqrt{3}}{4\pi}\approx 0.138\)\footnote[3]{References \cite{cardy, yu07, berg,kovacs}}&0.138(3)\footnote[4]{Reference \cite{kovacsPotts}} &G\\[3 pt]
          3 & $s_cs_c$& \parbox[c]{2em}{\includegraphics[width=0.3in]{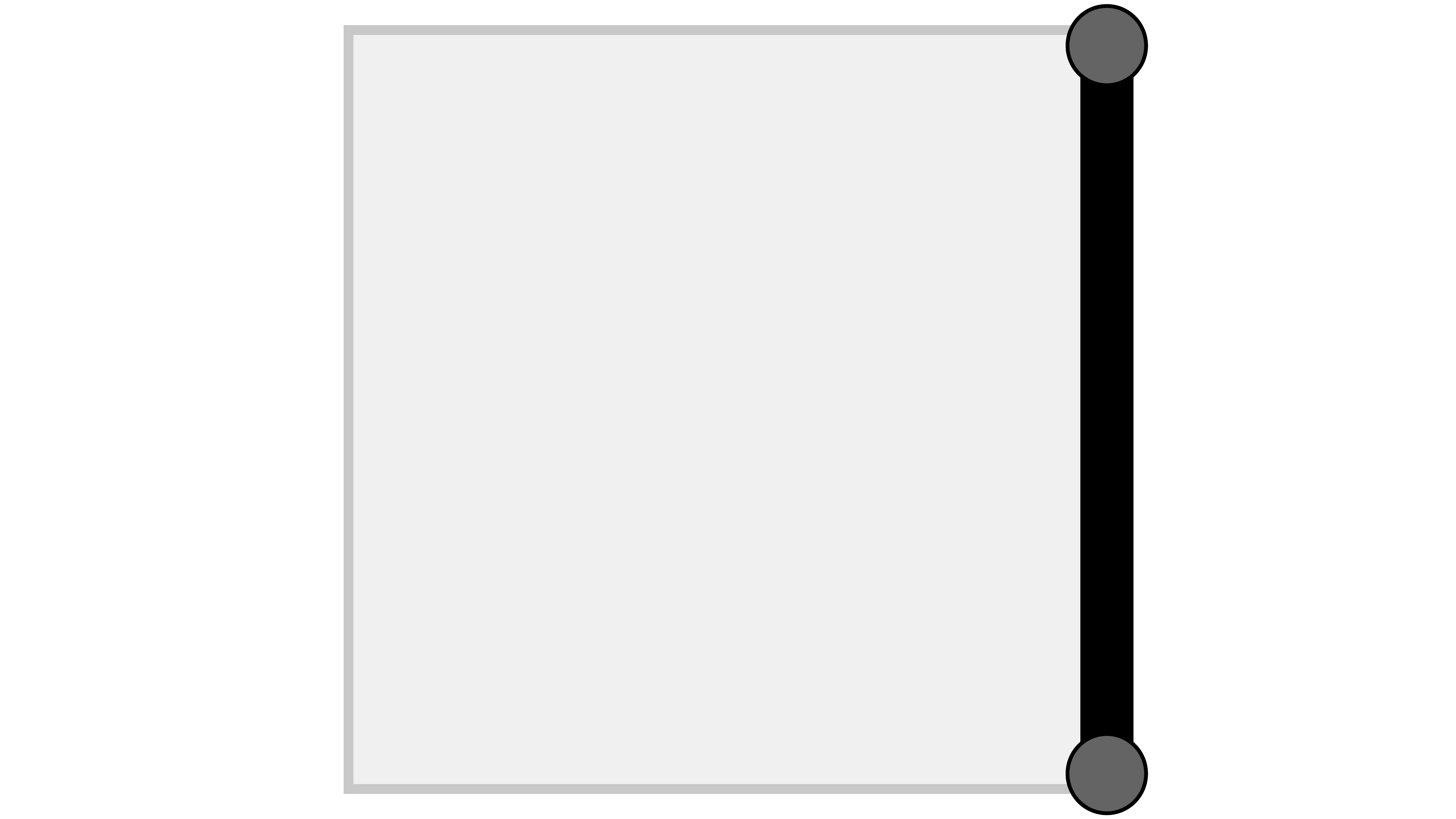}}  & $ 2S(\frac{\pi}{2})$& - & \(\frac{\sqrt{3}}{2\pi}\approx 0.276\)& 0.27(2) &BC\\[3 pt]
   4 & $ss_c$ & \parbox[c]{2em}{\includegraphics[width=0.3in]{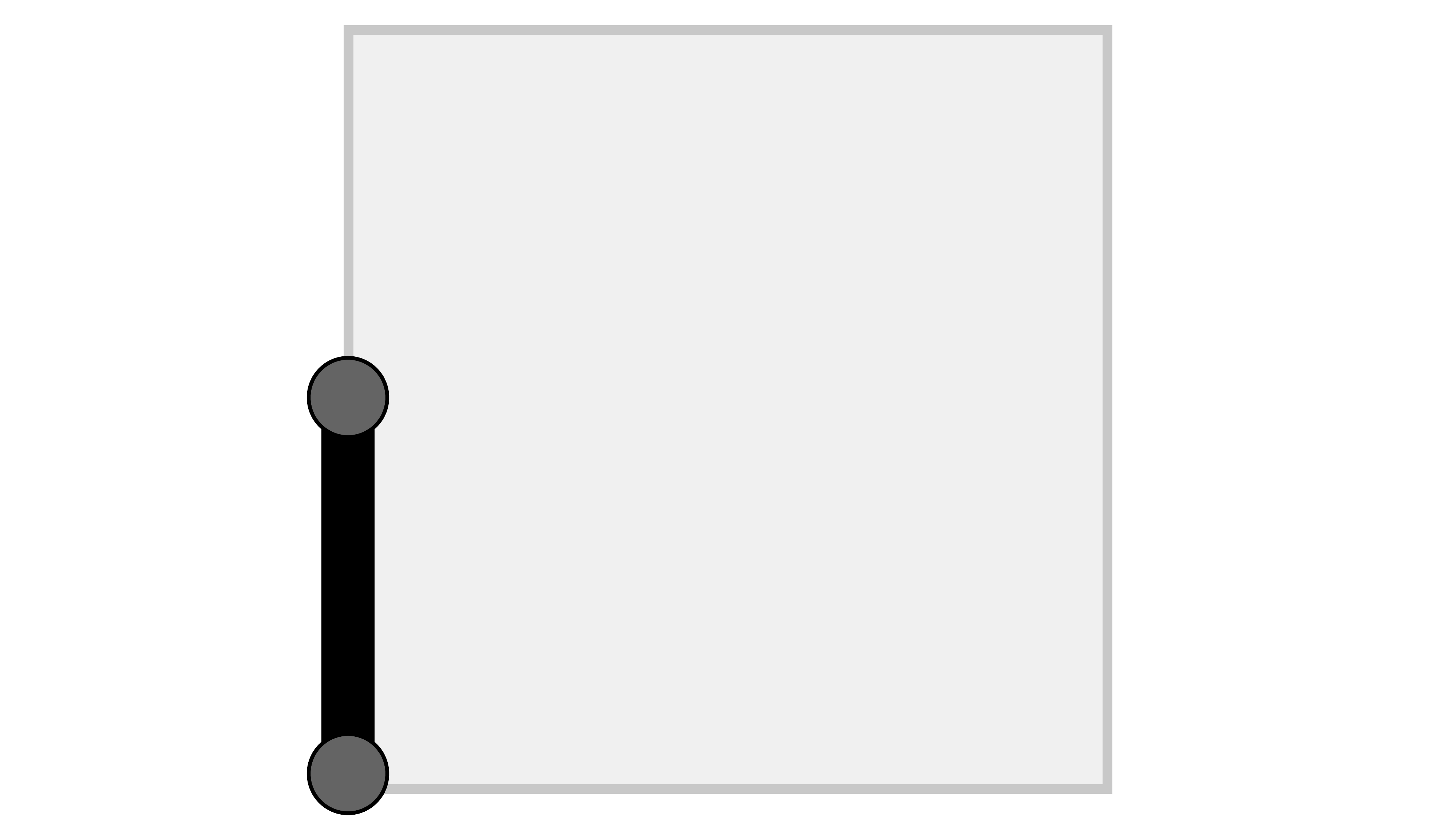}}  & $ S(\pi) + S(\frac{\pi}{2})$ & - & \(\frac{3\sqrt{3}}{8\pi}\approx 0.207\)& 0.21(1) &G \& BC \\[3 pt]
   5 & $t_c t_c$ &  \parbox[c]{2em}{\includegraphics[width=0.3in]{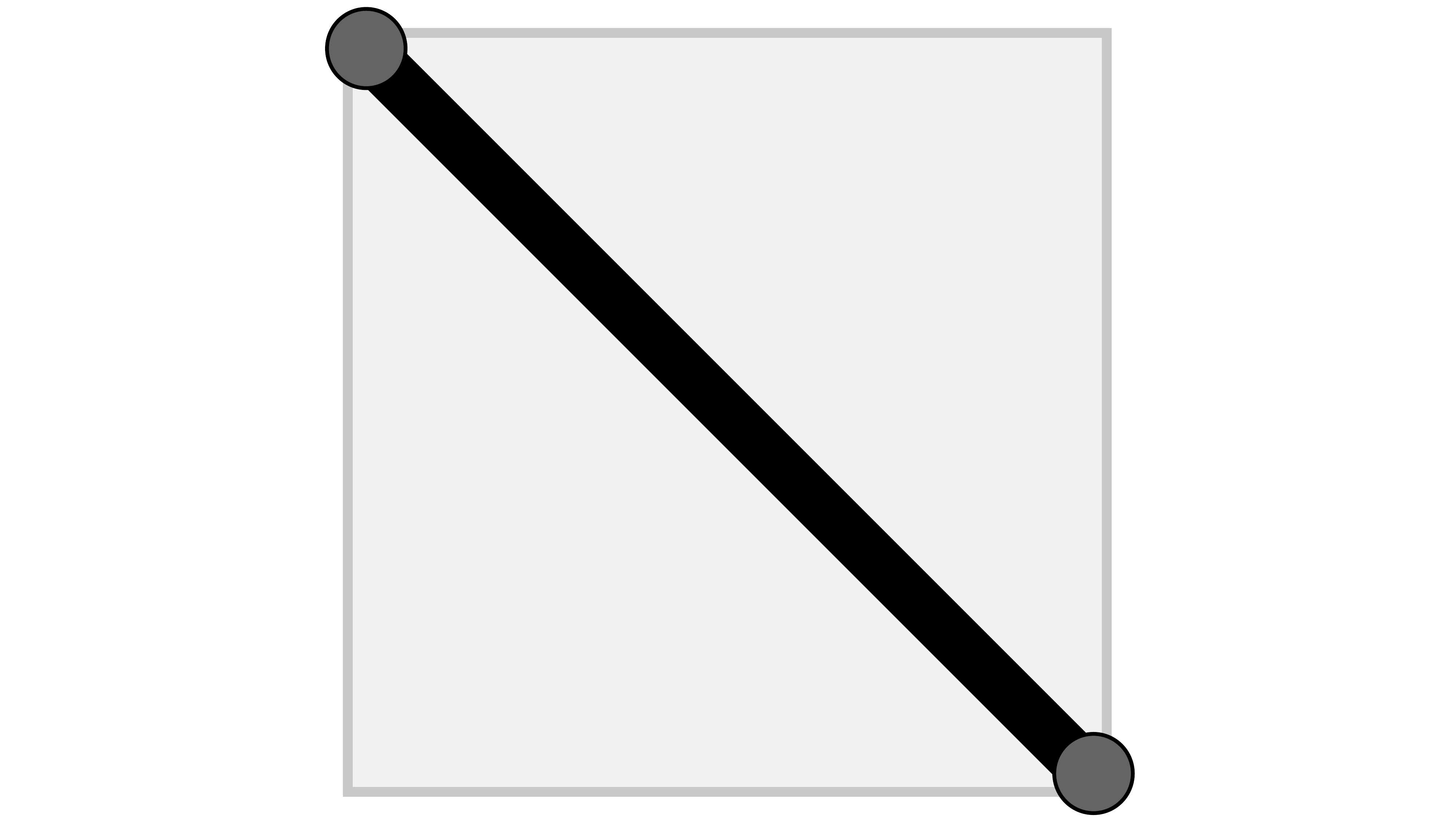}}   & $ 2T\left(\frac{\pi}{4},\frac{\pi}{2}\right)$& - & $\frac{11\sqrt{3}}{8\pi}\approx 0.758$& 0.78(2) & BC\\[3 pt]
     6 & $tt$ & \parbox[c]{2em}{\includegraphics[width=0.3in]{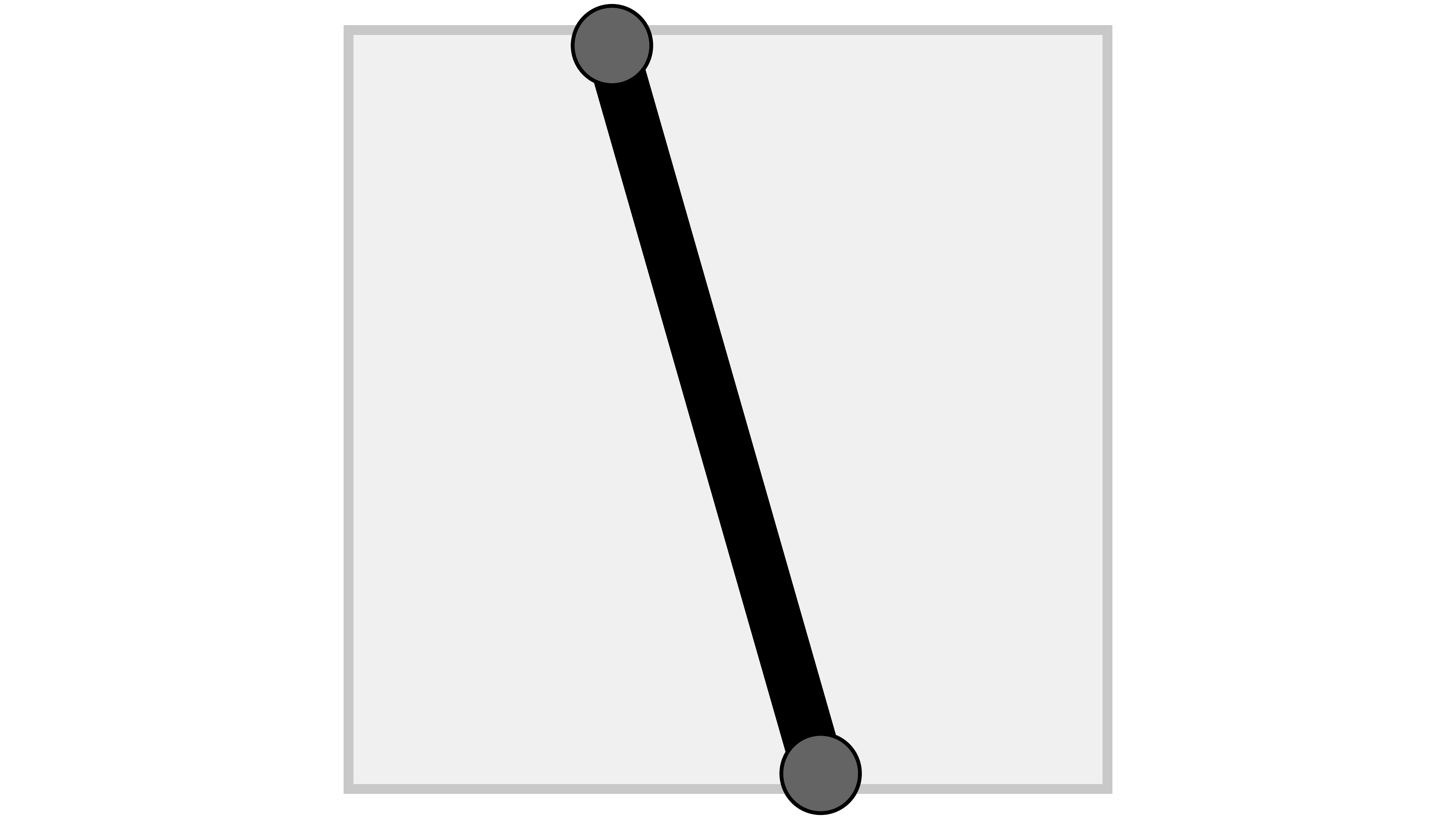}} & $ 2T(\gamma,\pi) $& $\frac{\pi}{2}$ & $\frac{11\sqrt{3}}{16\pi}\approx0.379$ & 0.38(1) & BC   \\[3 pt]
    7 & $tt'$  & \parbox[c]{2em}{\includegraphics[width=0.3in]{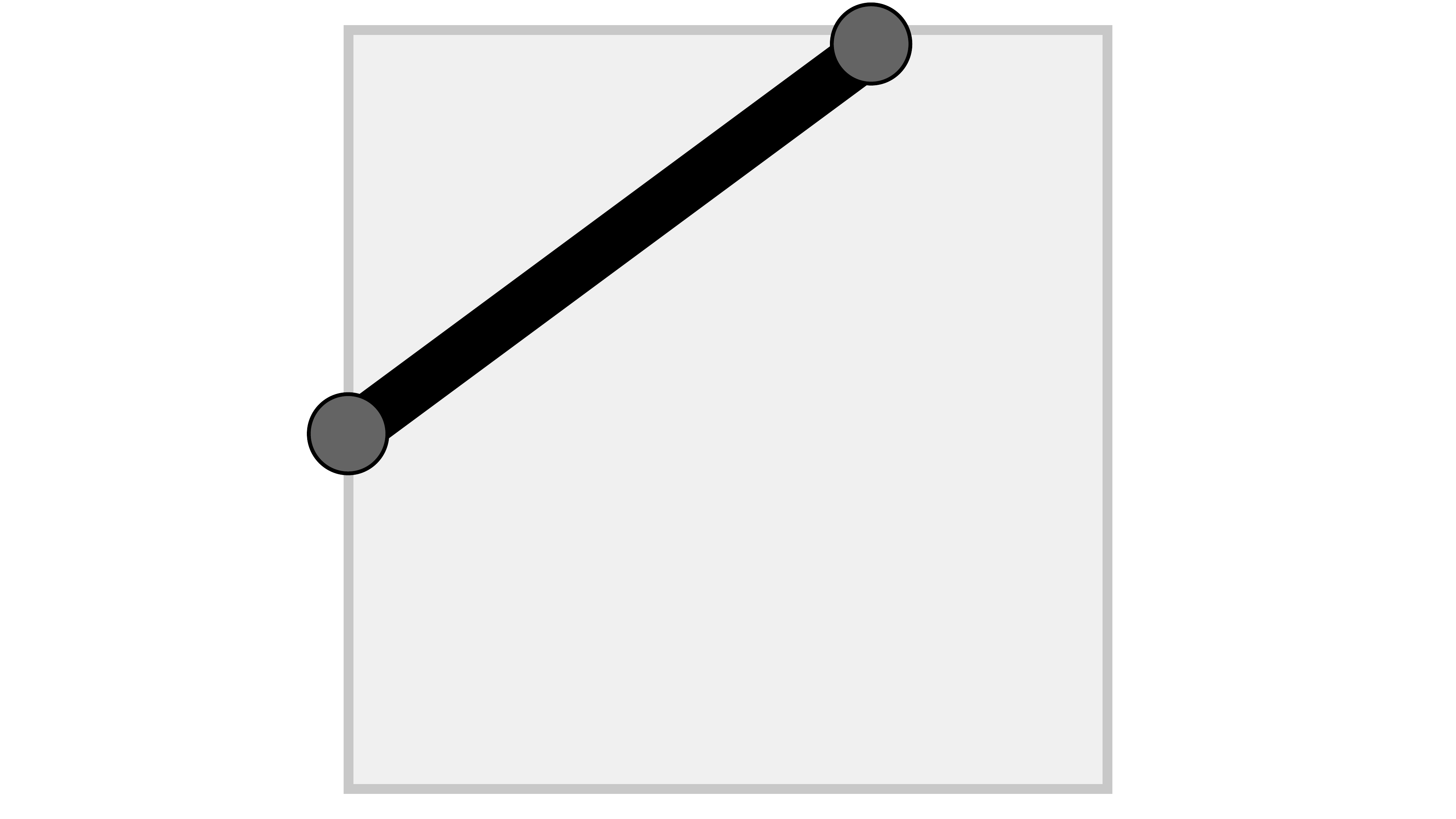}} & $T(\gamma,\pi)+T\left(\frac{\pi}{2}-\gamma,\pi\right) $ & $\frac{\pi}{4} $& $\frac{127\sqrt{3}}{144\pi}\approx0.486$ & 0.48(2)  &BC* \\[3 pt]
   8 & $tt_c$ &   \parbox[c]{2em}{\includegraphics[width=0.3in]{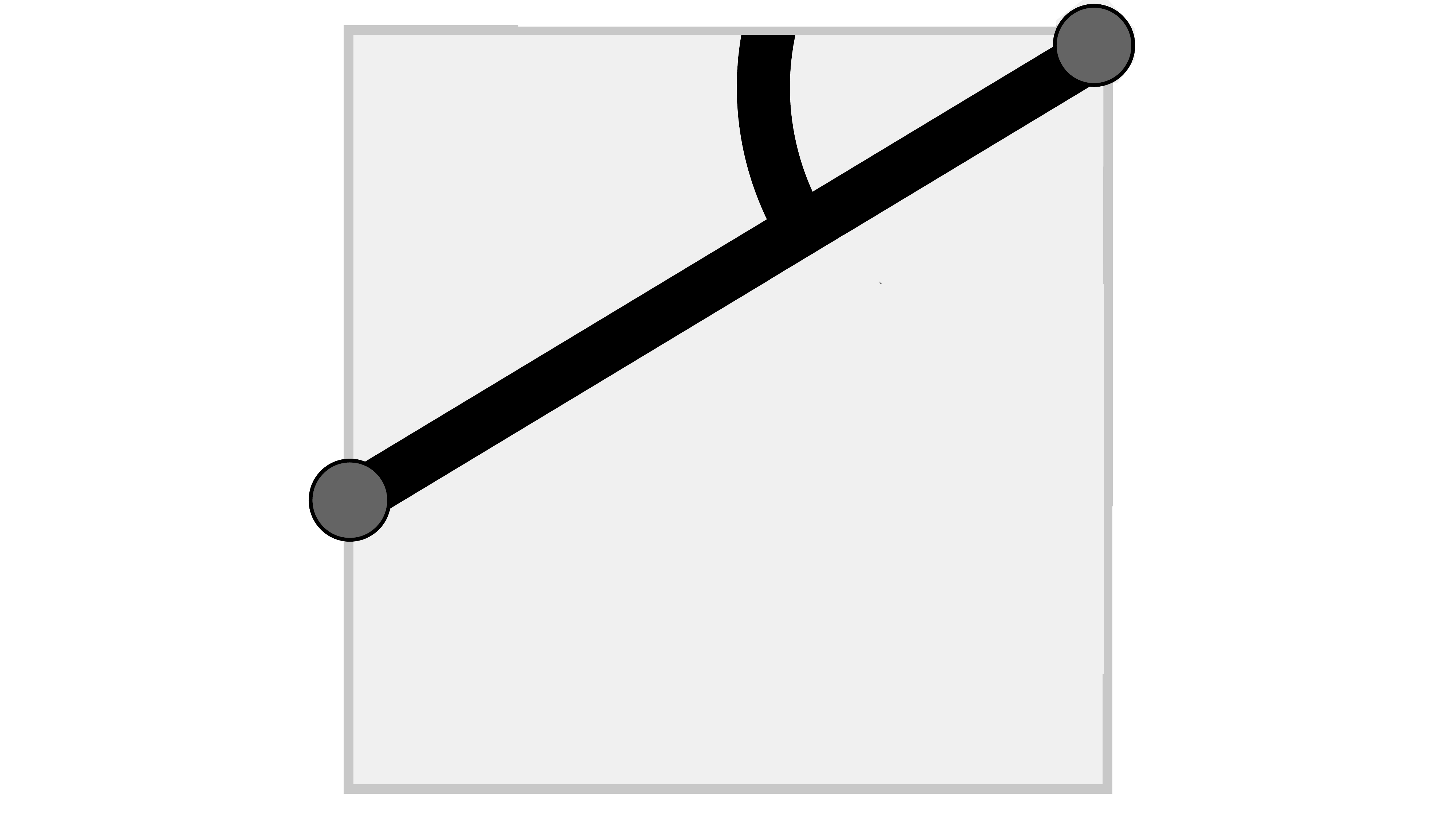}}  & $ T\left(\gamma,\frac{\pi}{2}\right)+T\left(\frac{\pi}{2}-\gamma,\pi\right)$& $ \arctan{\left(\frac{1}{2}\right)}$ & $\approx0.649$ &  0.66(3) &BC*\\[3 pt]
   9  & $tb$  &  \parbox[c]{2em}{\includegraphics[width=0.3in]{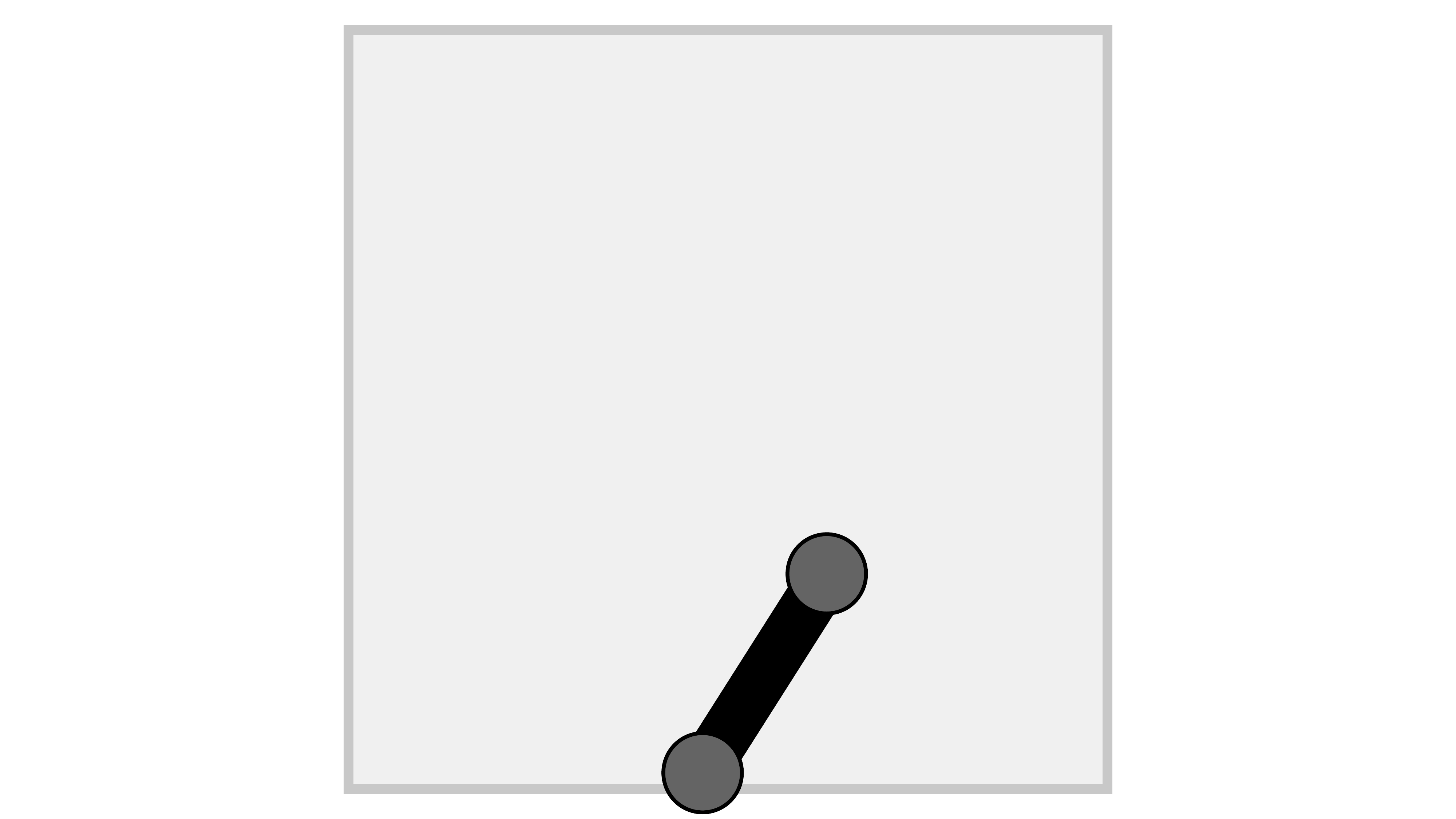}}   & $T(\gamma,\pi)+B(2\pi)$& $\frac{\pi}{2}$& $\frac{27\sqrt{3}}{64\pi}\approx0.233$ & 0.24(1) &G \& BC \\[3pt]
    10 & $t_cb$  & \parbox[c]{2em}{\includegraphics[width=0.3in]{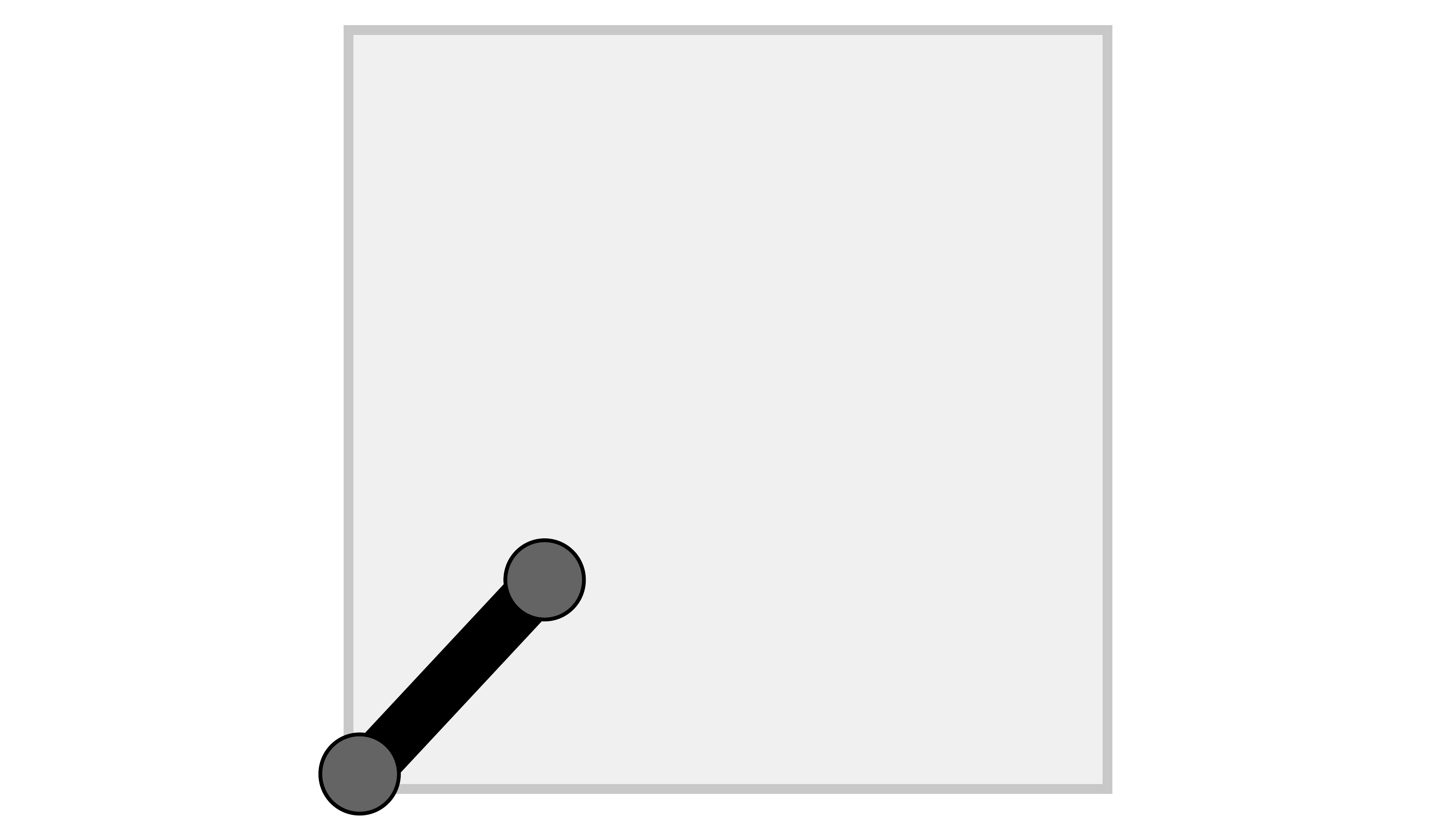}} & $ T\left(\gamma,\frac{\pi}{2}\right)+B(2\pi)$& $ \frac{\pi}{4}$ &$\frac{49\sqrt{3}}{64\pi}\approx0.422$ & 0.43(1)& G \& BC*\\
    \bottomrule
        \bottomrule
    \end{tabular}
    \label{tab:lines}
\end{table*}

\begin{figure}
    \centering
    \includegraphics[width=\linewidth]{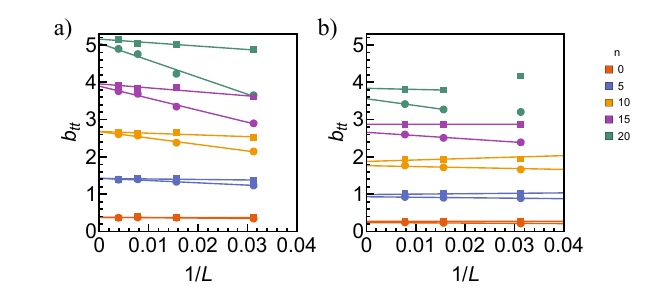} 
    \vskip-0.5cm
    \caption{Extrapolation of the cluster count exponent $b_{tt}$ for angles \(\gamma = \arctan{(1/n)} \) for a line segment spanning the entire system (Fig.~\ref{fig:ill}, line 6). As an illustration of universality, the results are shown for both site (\drawcircle[black]) and bond (\drawsquare[black]) percolation in (a) 2$d$ and (b) 3$d$.
    }
    \label{fig:bvs1overL}
\end{figure}

In our large-scale Monte Carlo calculations, we investigate cluster tomography for critical site and bond percolation 
on the simple cubic lattice in 2$d$ and 3$d$. 
The respective critical occupancies for site and percolation are \(p_c = 0.592746\) and \(p_c = 0.5\) in 2$d$ \cite{stauffer}, and $p_c=0.311608$ and $p_c=0.2488126$ in 3$d$ \cite{perc_cr,*perc_cr2,note1}.
We studied systems up to linear size $L=512$ in 2$d$ and $L=256$ in 3$d$, with at least $10\,000$ samples in each case.
For each line configuration, we calculate finite-size estimates of $b(L)$ through two-point fits at sizes $L$ and $L/2$.
We then estimate $b$ using a linear extrapolation of the largest sizes against $1/L$, 
as shown in Fig.~\ref{fig:bvs1overL} for a line segment spanning the entire system (line 6) in 2$d$ and 3$d$. 
The extrapolated $b$ values for site and bond percolation 
agree within the error, indicating universality. 
Therefore, all quoted numerical $b$ values are averaged for site and bond percolation.

\begin{figure}
    \centering
    \includegraphics[width=0.95\linewidth]{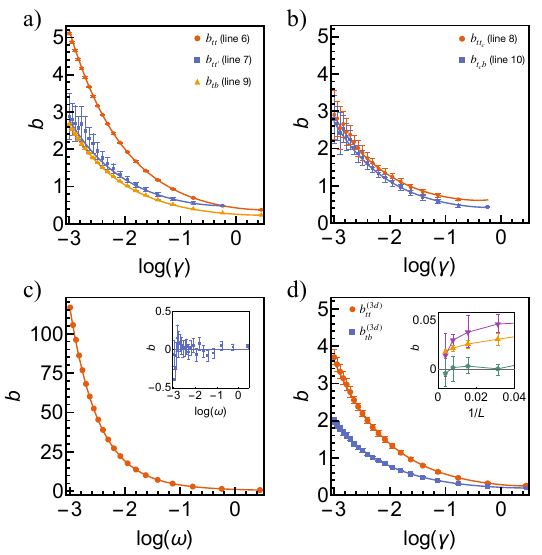} 
    \vskip-0.5cm
	\caption{(a-b) Angle dependence of the cluster count exponent $b$ in 2$d$ for lines 6-10 in Fig.~\ref{fig:ill}a. Numerically determined points 
 and curves showing the analytic predictions given in Table~\ref{tab:lines} are plotted for 
 (a) $b_{tt}$, 
  $b_{tt'}$ 
  and $b_{tb}$, 
   and 
 (b) $b_{tt_c}$ 
 and $b_{t_cb}$. 
 (c) $b_{t_ct_c}$ (line 5) in a sheared square system with acute angle $\omega$.
 (inset) The value of $b$ for a contour around the perimeter of a sheared square is zero.
 (d) Angle dependence of $b$ in 3$d$ for $b_{tt}^{(3d)}$ 
 and $b_{tb}^{(3d)}$. 
 Plotted lines show a fit of the form given in Eq.~(\ref{eq:b-fit-form}).
 (inset) Estimates of $b$ along a partial line segment on the surface of a cube (orange up triangles), 
 a full line segment through the center of the face with $\gamma=\pi/2$ (purple down triangles), 
 and a full line along the edge of a cube (green pentagons) 
 all tend to zero at large $L$.
}
    \label{fig:bvsgamma}
\end{figure}

In a 2$d$ square system, line types 1--5 each have universal $b$ values with no angle-dependence. 
Numerical estimates for lines 3--5 are listed in Table~\ref{tab:lines} along with known results for lines 1--2  \cite{cardy,yu07,kovacs,berg,kovacsPotts}.
The angle dependence of lines 6--10 can be explored to high precision for angles $\gamma = \arctan(1/n)$ for integer $n_{min}\leq n\leq 20$, where $n_{min}\to 0$ for lines 6 and 9, $n_{min} = 1$ for lines 7 and 10, and $n_{min}  = 2$ for line 8. 
For line 6, we can directly access the $\gamma$ dependence of $b$ to high precision using the boundary changing method by starting with FBC on one opposite pair of system edges and PBC on the other. 
This approach can be used in conjunction with the geometric method for line 9.
For lines 7, 8 and 10, the initial configuration must have FBC on all sides.
In these cases, $b$ values can be determined numerically to high precision by incorporating lines of type 6 and 9 into the initial configuration so that a closed loop forms after applying PBC, then subtracting their contribution from the count to determine the corner contribution of interest, 
see Fig. S1 of the Supplemental Material \cite{SI}.
Numerical results for lines 6-10 are plotted in Fig.~\ref{fig:bvsgamma}a-b, along with the predicted analytic form of the $\gamma$ dependence, and numerical values for $\gamma(n_{min})$ 
for each line type are listed in Table~\ref{tab:lines}.
In all cases, we observe a good agreement between numerical values and analytic predictions.

Due to the $-B(\omega)$ term in $T(\gamma,\omega)$ contributing only for type $t_c$ endpoints, we further test for the presence of this term by examining $b_{t_ct_c}$ for line 5 as the square system is sheared with acute angle $\omega$.
Analytically, we expect $b_{t_ct_c} = 2T(\gamma,\omega)$ with $\tan(\gamma) = \tan{\omega}/(1+\tan{\omega})$.
Numerically, this can be tested for $\tan{\omega} = 1/n$ with $\tan{\gamma} = 1/(n+1)$ in a system of size $L$ by $L+n$.
The results, plotted in Fig.~\ref{fig:bvsgamma}c, show good agreement between predictions and numerical values.
As another test of our analytic predictions, we also consider the cluster count around the entire perimeter of the sheared square system. 
Since there are no bulk corners and no boundary changes in this case, the corner contribution is expected to be zero, as demonstrated numerically in the inset of Fig.~\ref{fig:bvsgamma}c.

In a 3$d$ cubic system, there are 23 types of line segment configurations depending on whether the endpoints are touching faces, edges, corners, or the bulk, 
as illustrated in Fig.~S2 of the Supplemental Material \cite{SI}.
Of these, only the case where both endpoints of the line segment are in the bulk has been studied numerically, giving $b_{bb}^{(3d)} = 0.130(3)$ \cite{kovacs3dperc}.
Although there are no known analytic predictions in 3$d$, endpoints are still expected to contribute to $b$ independently. Therefore relations such as
\begin{equation}
	b_{tb} = (b_{tt} + b_{bb})/2
	\label{eq:mixed}
\end{equation}
will still hold true in 3$d$ percolation and other clustered systems.
We explore this relation by examining 
the angle dependence of a line segment spanning the cube and ending on opposite faces ($b_{tt}^{(3d)}$), and the case where the line segment starts on a face and ends in the bulk  ($b_{tb}^{(3d)}$), as shown in Fig.~\ref{fig:bvsgamma}d. Motivated by the expressions for $b_{tt}$ and $b_{tb}$ in 2$d$, we fit a line of the form
\begin{equation}
    b^{(3d)}(\gamma) = c_1\left(\frac{\pi}{\gamma}+\frac{\pi}{\pi-\gamma}\right) + c_2,
    \label{eq:b-fit-form}
\end{equation} 
which gives $c_1\sim0.06$ and $c_2\sim0.03$ for $b_{tt}^{(3d)}$ and $c_1\sim 0.03$ and $c_2\sim0.08$ for $b_{tb}^{(3d)}$.
Equation~(\ref{eq:mixed}) is satisfied within the numerical uncertainty on the data points. 
However, some line types might not to lead to a finite $b$ in 3$d$. 
In particular, the critical 3d percolation occupancy is below that in 2d at which the surface alone would lead to a logarithmic corner contribution.
We investigate three cases of a line segment on a surface of the cube: a partial line segment in a face, a full line segment along an edge, and a full line segment through the center of the face parallel to an edge. 
As shown in the inset of Fig.~\ref{fig:bvsgamma}d, these cases suggest 
that the cluster count exponent vanishes in 3$d$ for lines fully on the surface (types 1-10 in Fig. S2 of the Supplemental Material~\cite{SI}).
In Appendix B and Fig.~S3 of the Supplemental Material \cite{SI}, we discuss how this finding is related to the shape of the clusters. 

The presence of a corner contribution term at criticality provides a simple and effective way of pinpointing the critical point and the corresponding universality class of a clustered system. 
Cluster tomography with traversing endpoints is especially promising in this regard as it leads to the strongest signals in both 2$d$ and 3$d$, even more so at sharper angles. 
Although our results are presented for percolation, which is the $Q \to 1$ limit of the $Q$-state Potts model, similar logarithmic terms are expected to exist for other (not necessary integer) values of $Q$, such as for the Ising model with $Q=2$, as long as the transition is second order \cite{kovacs14}. 
Techniques from conformal field theory are expected to give the correct analytic formulas also in these cases, at least for Fortuin-Kasteleyn clusters \cite{Fortuin-Kasteleyn}, as explored in 2$d$ for lines 1 and 2 in Ref. \cite{kovacs14}.

Our results readily provide the universal term of skeletal entanglement in the critical bond-diluted transverse-field Ising model \cite{Kovacs2012gap, kovacs3dperc}, with 8 line types studied here for the first time in 2$d$. In this application, the natural quantity to study is the number of \emph{crossed} clusters, where only those clusters contribute that are not fully contained by the line. Our numerical findings confirm the expectation that this difference is irrelevant at the critical point, yielding the same asymptotic behavior in the quantum model. 


It would also be interesting to explore cluster tomography for percolation in higher dimensions, as well as in the mean-field limit, although there are clearly no corner contributions on the Bethe lattice.
In general, for higher dimensional systems, or systems lacking conformal invariance, the angle-dependence could be nontrivial, meaning that different configurations of cluster tomography could unveil distinct universal information.

In some cases, including experimental realizations, it may not be possible to implement changes in the boundary conditions, meaning that one or both numerical cancellation techniques may not apply.
However, in many such cases the nonlinear correction may still be measured to high precision while in other cases the nonlinear term can be accessed through comparison of direct measurements of $N(\ell)$ at different system sizes, leading to cancellation of the linear term on average. 
In Fig.~S4 of the Supplemental Material \cite{SI} we demonstrate that the presence of a nonlinear corner contribution and the corresponding $b$ exponent can be determined even without the advanced cancellation techniques, suggesting applicability of this methodology in a broad range of cluster-based systems. \\

%
%
%
We thank W.~Witczak-Krempa 
for helpful discussions. We would like to acknowledge the WCAS Summer Grant Award from the Weinberg College Baker Program in Undergraduate Research at Northwestern University. This work was supported by the National Science Foundation under Grant No.~PHY-2310706 of the QIS program in the Division of Physics. This research was supported in part through the computational resources and staff contributions provided for the Quest high performance computing facility at Northwestern University which is jointly supported by the Office of the Provost, the Office for Research, and Northwestern University Information Technology.

\appendix
\section{Analyic arguments}
Bond percolation can be considered as the $Q \to 1$ limit of the $Q$-state Potts model \cite{wu}, defined on a lattice with sites $i=1,2,\dots,n$ and $m$ nearest neighbor bonds. The partition sum of the Potts model is given by
\be
Z(Q)=\sum_{s} \prod_{\left\langle ij\right\rangle } \exp \left( K \delta_{s_i,s_j} \right)\;,
\label{Potts}
\ee
where $\delta_{s_i,s_j}$ is the Kronecker symbol and $K$ is the reduced coupling, which is the ratio of the pair interaction and the temperature.
Using the identity $\exp\left( K \delta_{s_i,s_j} \right)=1+\frac{p}{1-p} \delta_{s_i,s_j}$
with $p=1-e^{-K}$, the sum of products in $Z$ is written in terms of the so-called Fortuin-Kasteleyn clusters \cite{Fortuin-Kasteleyn}, denoted by $F$. In $F$ an edge of the lattice  $i,j$ is occupied if a factor $\frac{p}{1-p} \delta_{s_i,s_j}$ is present and
if the spins exist in the same state within any connected cluster. 
%
Up to an irrelevant prefactor, this leads to 
\be
Z(Q) \sim \sum_{F} Q^{N_{\mathrm{tot}}(F)} p^{M(F)} {(1-p)}^{m-M(F)}\;.
\label{Z}
\ee
For a given element of $F$ there are $N_{\mathrm{tot}}(F) \le n$ connected components and $M(F) \le m$ occupied bonds.
The mean total number of clusters is then given by
\be
\left\langle N_{\mathrm{tot}}\right\rangle= Q\dfrac{\partial \ln Z(Q)}{\partial Q}\;.
\label{N_tot}
\ee
If we fix all spins on a one-dimensional contour $\Gamma$ (in state $1$, say), but leave the couplings unchanged, this relation is modified as 
\be
\left\langle N_\mathrm{tot} - N_{\Gamma} \right\rangle=
Q\dfrac{\partial \ln Z_{\Gamma}(Q)}{\partial Q} \;.
\label{N_tot-N_Gamma}
\ee
At the critical point, $p=p_c$, we can write \cite{kovacs}
\be
\ln Z(Q)-\ln Z_{\Gamma}(Q)\sim  f_{\mathrm e}(Q)L_{\Gamma}  +f_{\mathrm c}(Q),
\label{log_Z}
\ee
where $L_{\Gamma}$ is the linear extension of $\Gamma$, $f_{\mathrm e}$ is the edge free-energy density, which is a non-universal quantity, and 
the term $f_{\mathrm c}(Q)$ originates from the corners (or endpoints) of $\Gamma$ and is known as the \emph{corner contribution} to the free-energy, which is expected to be universal. 
The corner contribution to the free energy can be expressed as \cite{cardypeschel,CardyBC,stephan2013}
\be
f_{\mathrm c}(Q)= \bigg(c(Q)A_{\Gamma}+\frac{\pi}{\omega_\Gamma}h(Q)\bigg)\ln L_{\Gamma}\;.
\label{FreeCornerEnergy}
\ee
The first term is the Cardy-Peschel term \cite{cardypeschel}, which receives contributions from each $\gamma_k$ corner of $\Gamma$, considering both the interior and exterior sides of the contour. 
Here $c(Q)$ is the central charge of the $Q$-state Potts model, and 
$A_{\Gamma}$ is a purely geometric factor that is the same for all values of $Q$ and emerges from the angles $\gamma_k$ as 
\be
A_{\Gamma}= \sum_k A_{\gamma_k} = \sum_k\frac{1}{24}\left(\frac{\gamma_k}{\pi}-\frac{\pi}{\gamma_k}\right)\;.
\label{AGamma}
\ee
The second term in Eq.~(\ref{FreeCornerEnergy}) is present each time there are different boundary conditions along the contour on either side of the corner, such as a change between fixed and free boundary conditions when a measuring contour touches a free surface.
The term $h(Q)$ is the scaling dimension of the boundary condition changing operator and $\omega_\Gamma$ includes the corner angle(s) at the location of the boundary condition change.

In this letter, we focus on a line segment of length $L_{\Gamma}=\ell$ and determine the corner contribution to the cluster number count. 
Using Eqs.~\eqref{N_tot} and \eqref{N_tot-N_Gamma} we can write 
\be
N(\ell)=Q\dfrac{df_{\mathrm e}(Q)}{dQ}\ell+Q\dfrac{df_{\mathrm c}(Q)}{dQ}\equiv a\ell+b\ln{\ell} \;,
\label{N_Gamma}
\ee
with $b$ being the cluster count exponent.
Percolation corresponds to $Q\to1$, so using the parametrizations $c(k)=1-6/[k(k+1)]$ and $\sqrt{Q}=2\cos(\frac{\pi}{k+1})$ \cite{CG} gives $c'(1)=5\sqrt{3}/(4\pi)$, where prime notation indicates a derivative with respect to $Q$. Additionally, changing the boundary condition from fixed along the line segment to free along a boundary gives $h'(1)=\sqrt{3}/(8\pi)$ \cite{cardy}. 
%
We conjecture that $b$ receives contributions from linear combinations of $B(\gamma) = c'(1)A_{\gamma_k}$ and $S(\gamma) = h'(1)\pi/\gamma$, corresponding to Eqs.~(\ref{eq:B}) and~(\ref{eq:S}) of the main text, 
depending on the endpoint types (surface, bulk, or traversing) of the line segment of interest, as discussed in the main text.

\section{Relation between cluster shape and $b$ }
The gap-size statistics $n(s)$ considers the distance $s$ between successive occurrences of a given cluster along a line segment through a clustered system, quantifying the frequency with which a gap of size $s$ occurs. 
This quantity provides a measure of the concavity of the cluster, and captures pairwise and higher order correlations in the structure.
For line types 1 and 2, the gap-size statistics is related to the corner contribution along a line segment of length $\ell\leq L/2$ with PBC through \cite{EPL}
\begin{equation}
C(\ell) = \frac{1}{L}\sum_{i=1}^{\ell}\sum_{s=i}^{L/2} n(s).
\end{equation}
In all previously studied cases, at criticality $n(s)\sim s^{-\zeta}$ with $\zeta = 2$ \cite{EPL,kovacsPotts, kovacs3dperc}.
After approximating the sum as an integral, 
this results in the logarithmic corner contribution to the cluster number count in Eq.~(\ref{N}). 

Measures of cluster tomography on the surface of a 3$d$ cube at criticality suggest $b=0$, indicating that $n(s)$ might decay faster. 
Exploring $n(s)$ of line segments 
on the face of a cube for site and bond percolation, shown in Fig.~S3 of the Supplemental Information \cite{SI}, we find universal behavior with exponent $\zeta = 2.25(1)$.
This further confirms the vanishing cluster count exponent 
in this case and provides an example of a critical system for which $\zeta \neq 2$.

\bibliography{tomography-sources}

\begin{thebibliography}{41}%
\makeatletter
\providecommand \@ifxundefined [1]{%
 \@ifx{#1\undefined}
}%
\providecommand \@ifnum [1]{%
 \ifnum #1\expandafter \@firstoftwo
 \else \expandafter \@secondoftwo
 \fi
}%
\providecommand \@ifx [1]{%
 \ifx #1\expandafter \@firstoftwo
 \else \expandafter \@secondoftwo
 \fi
}%
\providecommand \natexlab [1]{#1}%
\providecommand \enquote  [1]{``#1''}%
\providecommand \bibnamefont  [1]{#1}%
\providecommand \bibfnamefont [1]{#1}%
\providecommand \citenamefont [1]{#1}%
\providecommand \href@noop [0]{\@secondoftwo}%
\providecommand \href [0]{\begingroup \@sanitize@url \@href}%
\providecommand \@href[1]{\@@startlink{#1}\@@href}%
\providecommand \@@href[1]{\endgroup#1\@@endlink}%
\providecommand \@sanitize@url [0]{\catcode `\\12\catcode `\$12\catcode
  `\&12\catcode `\#12\catcode `\^12\catcode `\_12\catcode `\%12\relax}%
\providecommand \@@startlink[1]{}%
\providecommand \@@endlink[0]{}%
\providecommand \url  [0]{\begingroup\@sanitize@url \@url }%
\providecommand \@url [1]{\endgroup\@href {#1}{\urlprefix }}%
\providecommand \urlprefix  [0]{URL }%
\providecommand \Eprint [0]{\href }%
\providecommand \doibase [0]{https://doi.org/}%
\providecommand \selectlanguage [0]{\@gobble}%
\providecommand \bibinfo  [0]{\@secondoftwo}%
\providecommand \bibfield  [0]{\@secondoftwo}%
\providecommand \translation [1]{[#1]}%
\providecommand \BibitemOpen [0]{}%
\providecommand \bibitemStop [0]{}%
\providecommand \bibitemNoStop [0]{.\EOS\space}%
\providecommand \EOS [0]{\spacefactor3000\relax}%
\providecommand \BibitemShut  [1]{\csname bibitem#1\endcsname}%
\let\auto@bib@innerbib\@empty
\bibitem [{\citenamefont {Komogortsev}\ \emph {et~al.}(2019)\citenamefont
  {Komogortsev}, \citenamefont {Iskhakov},\ and\ \citenamefont
  {Fel\^{a}k}}]{komogortsev2019}%
  \BibitemOpen
  \bibfield  {author} {\bibinfo {author} {\bibfnamefont {S.~V.}\ \bibnamefont
  {Komogortsev}}, \bibinfo {author} {\bibfnamefont {R.~S.}\ \bibnamefont
  {Iskhakov}},\ and\ \bibinfo {author} {\bibfnamefont {V.~A.}\ \bibnamefont
  {Fel\^{a}k}},\ }\bibfield  {title} {\bibinfo {title} {Fractal dimension
  effect on the magnetization curves of exchange-coupled clusters of magnetic
  nanoparticles},\ }\href {https://doi.org/10.1134/S1063776119040095}
  {\bibfield  {journal} {\bibinfo  {journal} {J. Exp. Theor. Phys.}\ }\textbf
  {\bibinfo {volume} {128}},\ \bibinfo {pages} {754} (\bibinfo {year}
  {2019})}\BibitemShut {NoStop}%
\bibitem [{\citenamefont {Tailleur}\ and\ \citenamefont
  {Cates}(2008)}]{tailleur2008}%
  \BibitemOpen
  \bibfield  {author} {\bibinfo {author} {\bibfnamefont {J.}~\bibnamefont
  {Tailleur}}\ and\ \bibinfo {author} {\bibfnamefont {M.~E.}\ \bibnamefont
  {Cates}},\ }\bibfield  {title} {\bibinfo {title} {Statistical mechanics of
  interacting run-and-tumble bacteria},\ }\href
  {https://doi.org/10.1103/PhysRevLett.100.218103} {\bibfield  {journal}
  {\bibinfo  {journal} {Phys. Rev. Lett.}\ }\textbf {\bibinfo {volume} {100}},\
  \bibinfo {pages} {218103} (\bibinfo {year} {2008})}\BibitemShut {NoStop}%
\bibitem [{\citenamefont {Buttinoni}\ \emph {et~al.}(2013)\citenamefont
  {Buttinoni}, \citenamefont {Bialk\'e}, \citenamefont {K\"ummel},
  \citenamefont {L\"owen}, \citenamefont {Bechinger},\ and\ \citenamefont
  {Speck}}]{Buttinoni2013}%
  \BibitemOpen
  \bibfield  {author} {\bibinfo {author} {\bibfnamefont {I.}~\bibnamefont
  {Buttinoni}}, \bibinfo {author} {\bibfnamefont {J.}~\bibnamefont {Bialk\'e}},
  \bibinfo {author} {\bibfnamefont {F.}~\bibnamefont {K\"ummel}}, \bibinfo
  {author} {\bibfnamefont {H.}~\bibnamefont {L\"owen}}, \bibinfo {author}
  {\bibfnamefont {C.}~\bibnamefont {Bechinger}},\ and\ \bibinfo {author}
  {\bibfnamefont {T.}~\bibnamefont {Speck}},\ }\bibfield  {title} {\bibinfo
  {title} {Dynamical clustering and phase separation in suspensions of
  self-propelled colloidal particles},\ }\href
  {https://doi.org/10.1103/PhysRevLett.110.238301} {\bibfield  {journal}
  {\bibinfo  {journal} {Phys. Rev. Lett.}\ }\textbf {\bibinfo {volume} {110}},\
  \bibinfo {pages} {238301} (\bibinfo {year} {2013})}\BibitemShut {NoStop}%
\bibitem [{\citenamefont {Palacci}\ \emph {et~al.}(2013)\citenamefont
  {Palacci}, \citenamefont {Sacanna}, \citenamefont {Steinberg}, \citenamefont
  {Pine},\ and\ \citenamefont {Chaikin}}]{Palacci2013}%
  \BibitemOpen
  \bibfield  {author} {\bibinfo {author} {\bibfnamefont {J.}~\bibnamefont
  {Palacci}}, \bibinfo {author} {\bibfnamefont {S.}~\bibnamefont {Sacanna}},
  \bibinfo {author} {\bibfnamefont {A.~P.}\ \bibnamefont {Steinberg}}, \bibinfo
  {author} {\bibfnamefont {D.~J.}\ \bibnamefont {Pine}},\ and\ \bibinfo
  {author} {\bibfnamefont {P.~M.}\ \bibnamefont {Chaikin}},\ }\bibfield
  {title} {\bibinfo {title} {Living crystals of light-activated colloidal
  surfers},\ }\href {https://doi.org/10.1126/science.1230020} {\bibfield
  {journal} {\bibinfo  {journal} {Science}\ }\textbf {\bibinfo {volume}
  {339}},\ \bibinfo {pages} {936} (\bibinfo {year} {2013})}\BibitemShut
  {NoStop}%
\bibitem [{\citenamefont {Be'er}\ \emph {et~al.}(2020)\citenamefont {Be'er},
  \citenamefont {Ilkanaiv}, \citenamefont {Gross}, \citenamefont {Kearns},
  \citenamefont {Heidenreich}, \citenamefont {Baer},\ and\ \citenamefont
  {Ariel}}]{be'er2020}%
  \BibitemOpen
  \bibfield  {author} {\bibinfo {author} {\bibfnamefont {A.}~\bibnamefont
  {Be'er}}, \bibinfo {author} {\bibfnamefont {B.}~\bibnamefont {Ilkanaiv}},
  \bibinfo {author} {\bibfnamefont {R.}~\bibnamefont {Gross}}, \bibinfo
  {author} {\bibfnamefont {D.~B.}\ \bibnamefont {Kearns}}, \bibinfo {author}
  {\bibfnamefont {S.}~\bibnamefont {Heidenreich}}, \bibinfo {author}
  {\bibfnamefont {M.}~\bibnamefont {Baer}},\ and\ \bibinfo {author}
  {\bibfnamefont {G.}~\bibnamefont {Ariel}},\ }\bibfield  {title} {\bibinfo
  {title} {A phase diagram for bacterial swarming},\ }\bibfield  {journal}
  {\bibinfo  {journal} {Commun. Phys.}\ }\textbf {\bibinfo {volume} {3}},\
  \href {https://doi.org/10.1038/s42005-020-0327-1} {10.1038/s42005-020-0327-1}
  (\bibinfo {year} {2020})\BibitemShut {NoStop}%
\bibitem [{\citenamefont {Swart}(2002)}]{swart2002}%
  \BibitemOpen
  \bibfield  {author} {\bibinfo {author} {\bibfnamefont {G.~W.}\ \bibnamefont
  {Swart}},\ }\bibfield  {title} {\bibinfo {title} {Activated leukocyte cell
  adhesion molecule ({CD166/ALCAM}): Developmental and mechanistic aspects of
  cell clustering and cell migration},\ }\href
  {https://doi.org/10.1078/0171-9335-00256} {\bibfield  {journal} {\bibinfo
  {journal} {Eur. J. Cell Biol.}\ }\textbf {\bibinfo {volume} {81}},\ \bibinfo
  {pages} {313} (\bibinfo {year} {2002})}\BibitemShut {NoStop}%
\bibitem [{\citenamefont {Guo}\ \emph {et~al.}(2002)\citenamefont {Guo},
  \citenamefont {Lee}, \citenamefont {Kamar}, \citenamefont {Akiyama},\ and\
  \citenamefont {Pierce}}]{guo2002}%
  \BibitemOpen
  \bibfield  {author} {\bibinfo {author} {\bibfnamefont {H.}~\bibnamefont
  {Guo}}, \bibinfo {author} {\bibfnamefont {I.}~\bibnamefont {Lee}}, \bibinfo
  {author} {\bibfnamefont {M.}~\bibnamefont {Kamar}}, \bibinfo {author}
  {\bibfnamefont {S.~K.}\ \bibnamefont {Akiyama}},\ and\ \bibinfo {author}
  {\bibfnamefont {M.}~\bibnamefont {Pierce}},\ }\bibfield  {title} {\bibinfo
  {title} {Aberrant {N}-glycosylation of $\beta$1 integrin causes reduced
  $\alpha$5$\beta$1 integrin clustering and stimulates cell migration},\
  }\href@noop {} {\bibfield  {journal} {\bibinfo  {journal} {Cancer Res.}\
  }\textbf {\bibinfo {volume} {62}},\ \bibinfo {pages} {6837} (\bibinfo {year}
  {2002})}\BibitemShut {NoStop}%
\bibitem [{\citenamefont {Vicsek}\ \emph {et~al.}(1995)\citenamefont {Vicsek},
  \citenamefont {Czir\'ok}, \citenamefont {Ben-Jacob}, \citenamefont {Cohen},\
  and\ \citenamefont {Shochet}}]{vicsek1995}%
  \BibitemOpen
  \bibfield  {author} {\bibinfo {author} {\bibfnamefont {T.}~\bibnamefont
  {Vicsek}}, \bibinfo {author} {\bibfnamefont {A.}~\bibnamefont {Czir\'ok}},
  \bibinfo {author} {\bibfnamefont {E.}~\bibnamefont {Ben-Jacob}}, \bibinfo
  {author} {\bibfnamefont {I.}~\bibnamefont {Cohen}},\ and\ \bibinfo {author}
  {\bibfnamefont {O.}~\bibnamefont {Shochet}},\ }\bibfield  {title} {\bibinfo
  {title} {Novel type of phase transition in a system of self-driven
  particles},\ }\href {https://doi.org/10.1103/PhysRevLett.75.1226} {\bibfield
  {journal} {\bibinfo  {journal} {Phys. Rev. Lett.}\ }\textbf {\bibinfo
  {volume} {75}},\ \bibinfo {pages} {1226} (\bibinfo {year}
  {1995})}\BibitemShut {NoStop}%
\bibitem [{\citenamefont {Cavagna}\ and\ \citenamefont
  {Giardina}(2014)}]{cavagna2014}%
  \BibitemOpen
  \bibfield  {author} {\bibinfo {author} {\bibfnamefont {A.}~\bibnamefont
  {Cavagna}}\ and\ \bibinfo {author} {\bibfnamefont {I.}~\bibnamefont
  {Giardina}},\ }\bibfield  {title} {\bibinfo {title} {Bird flocks as condensed
  matter},\ }\href {https://doi.org/10.1146/annurev-conmatphys-031113-133834}
  {\bibfield  {journal} {\bibinfo  {journal} {Annu. Rev. Condens. Matter
  Phys.}\ }\textbf {\bibinfo {volume} {5}},\ \bibinfo {pages} {183} (\bibinfo
  {year} {2014})}\BibitemShut {NoStop}%
\bibitem [{\citenamefont {Katz}\ \emph {et~al.}(2011)\citenamefont {Katz},
  \citenamefont {Tunstr{\o}m}, \citenamefont {Ioannou}, \citenamefont {Huepe},\
  and\ \citenamefont {Couzin}}]{katz2011}%
  \BibitemOpen
  \bibfield  {author} {\bibinfo {author} {\bibfnamefont {Y.}~\bibnamefont
  {Katz}}, \bibinfo {author} {\bibfnamefont {K.}~\bibnamefont {Tunstr{\o}m}},
  \bibinfo {author} {\bibfnamefont {C.~C.}\ \bibnamefont {Ioannou}}, \bibinfo
  {author} {\bibfnamefont {C.}~\bibnamefont {Huepe}},\ and\ \bibinfo {author}
  {\bibfnamefont {I.~D.}\ \bibnamefont {Couzin}},\ }\bibfield  {title}
  {\bibinfo {title} {Inferring the structure and dynamics of interactions in
  schooling fish},\ }\href {https://doi.org/10.1073/pnas.1107583108} {\bibfield
   {journal} {\bibinfo  {journal} {Proc. Natl. Acad. Sci.}\ }\textbf {\bibinfo
  {volume} {108}},\ \bibinfo {pages} {18720} (\bibinfo {year}
  {2011})}\BibitemShut {NoStop}%
\bibitem [{\citenamefont {Gordon}(2014)}]{gordon2014}%
  \BibitemOpen
  \bibfield  {author} {\bibinfo {author} {\bibfnamefont {D.~M.}\ \bibnamefont
  {Gordon}},\ }\bibfield  {title} {\bibinfo {title} {The ecology of collective
  behavior},\ }\href {https://doi.org/10.1371/journal.pbio.1001805} {\bibfield
  {journal} {\bibinfo  {journal} {PLOS Biology}\ }\textbf {\bibinfo {volume}
  {12}},\ \bibinfo {pages} {e1001805} (\bibinfo {year} {2014})}\BibitemShut
  {NoStop}%
\bibitem [{\citenamefont {Kov\'acs}\ \emph
  {et~al.}(2014{\natexlab{a}})\citenamefont {Kov\'acs}, \citenamefont
  {El\c{c}i}, \citenamefont {Weigel},\ and\ \citenamefont
  {Igl\'oi}}]{kovacs14}%
  \BibitemOpen
  \bibfield  {author} {\bibinfo {author} {\bibfnamefont {I.~A.}\ \bibnamefont
  {Kov\'acs}}, \bibinfo {author} {\bibfnamefont {E.~M.}\ \bibnamefont
  {El\c{c}i}}, \bibinfo {author} {\bibfnamefont {M.}~\bibnamefont {Weigel}},\
  and\ \bibinfo {author} {\bibfnamefont {F.}~\bibnamefont {Igl\'oi}},\
  }\bibfield  {title} {\bibinfo {title} {Corner contribution to cluster numbers
  in the {P}otts model.},\ }\href {https://doi.org/10.1103/PhysRevB.89.064421}
  {\bibfield  {journal} {\bibinfo  {journal} {Phys. Rev. B}\ }\textbf {\bibinfo
  {volume} {89}},\ \bibinfo {pages} {064421} (\bibinfo {year}
  {2014}{\natexlab{a}})}\BibitemShut {NoStop}%
\bibitem [{\citenamefont {Stauffer}\ and\ \citenamefont
  {Aharony}(1992)}]{stauffer}%
  \BibitemOpen
  \bibfield  {author} {\bibinfo {author} {\bibfnamefont {D.}~\bibnamefont
  {Stauffer}}\ and\ \bibinfo {author} {\bibfnamefont {A.}~\bibnamefont
  {Aharony}},\ }\href@noop {} {\emph {\bibinfo {title} {Introduction to
  percolation theory}}}\ (\bibinfo  {publisher} {Taylor \& Francis},\ \bibinfo
  {address} {London},\ \bibinfo {year} {1992})\BibitemShut {NoStop}%
\bibitem [{\citenamefont {Smirnov}(2001)}]{smirnov}%
  \BibitemOpen
  \bibfield  {author} {\bibinfo {author} {\bibfnamefont {S.}~\bibnamefont
  {Smirnov}},\ }\bibfield  {title} {\bibinfo {title} {Critical percolation in
  the plane: conformal invariance, {C}ardy's formula, scaling limits},\ }\href
  {https://doi.org/10.1016/S0764-4442(01)01991-7} {\bibfield  {journal}
  {\bibinfo  {journal} {C. R. Acad. Sci. Paris S\'er. I Math.}\ }\textbf
  {\bibinfo {volume} {333}},\ \bibinfo {pages} {239} (\bibinfo {year}
  {2001})}\BibitemShut {NoStop}%
\bibitem [{\citenamefont {Tsai}\ \emph {et~al.}(2013)\citenamefont {Tsai},
  \citenamefont {Yam},\ and\ \citenamefont {Zhou}}]{tsai2013}%
  \BibitemOpen
  \bibfield  {author} {\bibinfo {author} {\bibfnamefont {J.}~\bibnamefont
  {Tsai}}, \bibinfo {author} {\bibfnamefont {S.~C.~P.}\ \bibnamefont {Yam}},\
  and\ \bibinfo {author} {\bibfnamefont {W.}~\bibnamefont {Zhou}},\ }\href@noop
  {} {\bibinfo {title} {Conformal invariance of the exploration path in 2-d
  critical bond percolation in the square lattice}} (\bibinfo {year} {2013}),\
  \Eprint {https://arxiv.org/abs/1112.2017} {arXiv:1112.2017 [math.PR]}
  \BibitemShut {NoStop}%
\bibitem [{\citenamefont {Cardy}(1987)}]{conf_inv}%
  \BibitemOpen
  \bibfield  {author} {\bibinfo {author} {\bibfnamefont {J.~L.}\ \bibnamefont
  {Cardy}},\ }\bibfield  {title} {\bibinfo {title} {Conformal invariance},\
  }in\ \href@noop {} {\emph {\bibinfo {booktitle} {Phase Transitions and
  Critical Phenomena}}},\ Vol.~\bibinfo {volume} {11},\ \bibinfo {editor}
  {edited by\ \bibinfo {editor} {\bibfnamefont {C.}~\bibnamefont {Domb}}\ and\
  \bibinfo {editor} {\bibfnamefont {J.}~\bibnamefont {Lebowitz}}}\ (\bibinfo
  {publisher} {Academic Press},\ \bibinfo {address} {London},\ \bibinfo {year}
  {1987})\ p.~\bibinfo {pages} {55}\BibitemShut {NoStop}%
\bibitem [{\citenamefont {Schramm}(2000)}]{sle}%
  \BibitemOpen
  \bibfield  {author} {\bibinfo {author} {\bibfnamefont {O.}~\bibnamefont
  {Schramm}},\ }\bibfield  {title} {\bibinfo {title} {Scaling limits of
  loop-erased random walks and uniform spanning trees},\ }\href
  {https://doi.org/10.1007/BF02803524} {\bibfield  {journal} {\bibinfo
  {journal} {Israel J. Math.}\ }\textbf {\bibinfo {volume} {118}},\ \bibinfo
  {pages} {221} (\bibinfo {year} {2000})}\BibitemShut {NoStop}%
\bibitem [{\citenamefont {Smirnov}\ and\ \citenamefont
  {Werner}(2001)}]{SmirnovWerner2001}%
  \BibitemOpen
  \bibfield  {author} {\bibinfo {author} {\bibfnamefont {S.}~\bibnamefont
  {Smirnov}}\ and\ \bibinfo {author} {\bibfnamefont {W.}~\bibnamefont
  {Werner}},\ }\bibfield  {title} {\bibinfo {title} {Critical exponents for
  two-dimensional percolation},\ }\href@noop {} {\bibfield  {journal} {\bibinfo
   {journal} {Math. Res. Lett.}\ }\textbf {\bibinfo {volume} {8}},\ \bibinfo
  {pages} {729} (\bibinfo {year} {2001})}\BibitemShut {NoStop}%
\bibitem [{\citenamefont {Dotsenko}\ and\ \citenamefont
  {Fateev}(1984)}]{dotsenko_fateev}%
  \BibitemOpen
  \bibfield  {author} {\bibinfo {author} {\bibfnamefont {V.}~\bibnamefont
  {Dotsenko}}\ and\ \bibinfo {author} {\bibfnamefont {V.}~\bibnamefont
  {Fateev}},\ }\bibfield  {title} {\bibinfo {title} {Conformal algebra and
  multipoint correlation functions in 2{D} statistical models},\ }\href
  {https://doi.org/10.1016/0550-3213(84)90269-4} {\bibfield  {journal}
  {\bibinfo  {journal} {Nucl. Phys. B}\ }\textbf {\bibinfo {volume} {240}},\
  \bibinfo {pages} {312} (\bibinfo {year} {1984})}\BibitemShut {NoStop}%
\bibitem [{\citenamefont {Cardy}(1992)}]{crossing}%
  \BibitemOpen
  \bibfield  {author} {\bibinfo {author} {\bibfnamefont {J.~L.}\ \bibnamefont
  {Cardy}},\ }\bibfield  {title} {\bibinfo {title} {Critical percolation in
  finite geometries},\ }\href {https://doi.org/10.1088/0305-4470/25/4/009}
  {\bibfield  {journal} {\bibinfo  {journal} {J. Phys. A}\ }\textbf {\bibinfo
  {volume} {25}},\ \bibinfo {pages} {L201} (\bibinfo {year}
  {1992})}\BibitemShut {NoStop}%
\bibitem [{\citenamefont {Yu}\ \emph {et~al.}(2008)\citenamefont {Yu},
  \citenamefont {Saleur},\ and\ \citenamefont {Haas}}]{yu07}%
  \BibitemOpen
  \bibfield  {author} {\bibinfo {author} {\bibfnamefont {R.}~\bibnamefont
  {Yu}}, \bibinfo {author} {\bibfnamefont {H.}~\bibnamefont {Saleur}},\ and\
  \bibinfo {author} {\bibfnamefont {S.}~\bibnamefont {Haas}},\ }\bibfield
  {title} {\bibinfo {title} {Entanglement entropy in the two-dimensional random
  transverse field ising model},\ }\href
  {https://doi.org/10.1103/PhysRevB.77.140402} {\bibfield  {journal} {\bibinfo
  {journal} {Phys. Rev. B}\ }\textbf {\bibinfo {volume} {77}},\ \bibinfo
  {pages} {140402} (\bibinfo {year} {2008})}\BibitemShut {NoStop}%
\bibitem [{\citenamefont {Kov\'acs}\ \emph {et~al.}(2012)\citenamefont
  {Kov\'acs}, \citenamefont {Igl\'oi},\ and\ \citenamefont {Cardy}}]{kovacs}%
  \BibitemOpen
  \bibfield  {author} {\bibinfo {author} {\bibfnamefont {I.~A.}\ \bibnamefont
  {Kov\'acs}}, \bibinfo {author} {\bibfnamefont {F.}~\bibnamefont {Igl\'oi}},\
  and\ \bibinfo {author} {\bibfnamefont {J.}~\bibnamefont {Cardy}},\ }\bibfield
   {title} {\bibinfo {title} {Corner contribution to percolation cluster
  numbers.},\ }\href {https://doi.org/10.1103/PhysRevB.86.214203} {\bibfield
  {journal} {\bibinfo  {journal} {Phys. Rev. B}\ }\textbf {\bibinfo {volume}
  {86}},\ \bibinfo {pages} {214203} (\bibinfo {year} {2012})}\BibitemShut
  {NoStop}%
\bibitem [{\citenamefont {Berthiere}\ and\ \citenamefont
  {Witczak-Krempa}(2022)}]{skeletal}%
  \BibitemOpen
  \bibfield  {author} {\bibinfo {author} {\bibfnamefont {C.}~\bibnamefont
  {Berthiere}}\ and\ \bibinfo {author} {\bibfnamefont {W.}~\bibnamefont
  {Witczak-Krempa}},\ }\bibfield  {title} {\bibinfo {title} {Entanglement of
  skeletal regions},\ }\href {https://doi.org/10.1103/PhysRevLett.128.240502}
  {\bibfield  {journal} {\bibinfo  {journal} {Phys. Rev. Lett.}\ }\textbf
  {\bibinfo {volume} {128}},\ \bibinfo {pages} {240502} (\bibinfo {year}
  {2022})}\BibitemShut {NoStop}%
\bibitem [{\citenamefont {Kov\'acs}\ and\ \citenamefont
  {Igl\'oi}(2014)}]{kovacs3dperc}%
  \BibitemOpen
  \bibfield  {author} {\bibinfo {author} {\bibfnamefont {I.~A.}\ \bibnamefont
  {Kov\'acs}}\ and\ \bibinfo {author} {\bibfnamefont {F.}~\bibnamefont
  {Igl\'oi}},\ }\bibfield  {title} {\bibinfo {title} {Corner contribution to
  percolation cluster numbers in three dimensions.},\ }\href
  {https://doi.org/10.1103/PhysRevB.89.174202} {\bibfield  {journal} {\bibinfo
  {journal} {Phys. Rev. B}\ }\textbf {\bibinfo {volume} {89}},\ \bibinfo
  {pages} {174202} (\bibinfo {year} {2014})}\BibitemShut {NoStop}%
\bibitem [{\citenamefont {Kov\'acs}\ \emph
  {et~al.}(2014{\natexlab{b}})\citenamefont {Kov\'acs}, \citenamefont
  {d'Auriac},\ and\ \citenamefont {Igl\'oi}}]{kovacs_largeQ}%
  \BibitemOpen
  \bibfield  {author} {\bibinfo {author} {\bibfnamefont {I.~A.}\ \bibnamefont
  {Kov\'acs}}, \bibinfo {author} {\bibfnamefont {J.-C.~A.}\ \bibnamefont
  {d'Auriac}},\ and\ \bibinfo {author} {\bibfnamefont {F.}~\bibnamefont
  {Igl\'oi}},\ }\bibfield  {title} {\bibinfo {title} {Excess entropy and
  central charge of the two-dimensional random-bond potts model in the large-q
  limit.},\ }\href {https://doi.org/10.1088/1742-5468/2014/09/P09019}
  {\bibfield  {journal} {\bibinfo  {journal} {J. Stat. Mech. Theory Exp.}\
  }\textbf {\bibinfo {volume} {2014}},\ \bibinfo {pages} {P09019} (\bibinfo
  {year} {2014}{\natexlab{b}})}\BibitemShut {NoStop}%
\bibitem [{\citenamefont {Witczak-Krempa}(2019)}]{Witczak-Krempa2019}%
  \BibitemOpen
  \bibfield  {author} {\bibinfo {author} {\bibfnamefont {W.}~\bibnamefont
  {Witczak-Krempa}},\ }\bibfield  {title} {\bibinfo {title} {Entanglement
  susceptibilities and universal geometric entanglement entropy},\ }\href
  {https://doi.org/10.1103/PhysRevB.99.075138} {\bibfield  {journal} {\bibinfo
  {journal} {Phys. Rev. B}\ }\textbf {\bibinfo {volume} {99}},\ \bibinfo
  {pages} {075138} (\bibinfo {year} {2019})}\BibitemShut {NoStop}%
\bibitem [{\citenamefont {Cardy}\ and\ \citenamefont
  {Peschel}(1988)}]{cardypeschel}%
  \BibitemOpen
  \bibfield  {author} {\bibinfo {author} {\bibfnamefont {J.~L.}\ \bibnamefont
  {Cardy}}\ and\ \bibinfo {author} {\bibfnamefont {I.}~\bibnamefont
  {Peschel}},\ }\bibfield  {title} {\bibinfo {title} {Finite-size dependence of
  the free energy in two-dimensional critical systems},\ }\href
  {https://doi.org/10.1016/0550-3213(88)90604-9} {\bibfield  {journal}
  {\bibinfo  {journal} {Nuclear Physics B}\ }\textbf {\bibinfo {volume}
  {300}},\ \bibinfo {pages} {377} (\bibinfo {year} {1988})}\BibitemShut
  {NoStop}%
\bibitem [{\citenamefont {Cardy}(1989)}]{CardyBC}%
  \BibitemOpen
  \bibfield  {author} {\bibinfo {author} {\bibfnamefont {J.~L.}\ \bibnamefont
  {Cardy}},\ }\bibfield  {title} {\bibinfo {title} {Boundary conditions, fusion
  rules and the {V}erlinde formula},\ }\href
  {https://doi.org/10.1016/0550-3213(89)90521-X} {\bibfield  {journal}
  {\bibinfo  {journal} {Nuclear Physics B}\ }\textbf {\bibinfo {volume}
  {324}},\ \bibinfo {pages} {581} (\bibinfo {year} {1989})}\BibitemShut
  {NoStop}%
\bibitem [{\citenamefont {St\'{e}phan}\ and\ \citenamefont
  {Dubail}(2013)}]{stephan2013}%
  \BibitemOpen
  \bibfield  {author} {\bibinfo {author} {\bibfnamefont {J.-M.}\ \bibnamefont
  {St\'{e}phan}}\ and\ \bibinfo {author} {\bibfnamefont {J.}~\bibnamefont
  {Dubail}},\ }\bibfield  {title} {\bibinfo {title} {Logarithmic corrections to
  the free energy from sharp corners with angle 2$\pi$},\ }\href
  {https://doi.org/10.1088/1742-5468/2013/09/P09002} {\bibfield  {journal}
  {\bibinfo  {journal} {J. Stat. Mech}\ ,\ \bibinfo {pages} {P09002}} (\bibinfo
  {year} {2013})}\BibitemShut {NoStop}%
\bibitem [{\citenamefont {van~den Berg}\ and\ \citenamefont
  {Conijn}(2016)}]{berg}%
  \BibitemOpen
  \bibfield  {author} {\bibinfo {author} {\bibfnamefont {J.}~\bibnamefont
  {van~den Berg}}\ and\ \bibinfo {author} {\bibfnamefont {R.}~\bibnamefont
  {Conijn}},\ }\bibfield  {title} {\bibinfo {title} {{The expected number of
  critical percolation clusters intersecting a line segment}},\ }\href
  {https://doi.org/10.1214/16-ECP4452} {\bibfield  {journal} {\bibinfo
  {journal} {Electron. Commun. Probab.}\ }\textbf {\bibinfo {volume} {21}},\
  \bibinfo {pages} {1 } (\bibinfo {year} {2016})}\BibitemShut {NoStop}%
\bibitem [{\citenamefont {Cardy}(2001)}]{cardy}%
  \BibitemOpen
  \bibfield  {author} {\bibinfo {author} {\bibfnamefont {J.}~\bibnamefont
  {Cardy}},\ }\href@noop {} {\bibinfo {title} {Conformal invariance and
  percolation}} (\bibinfo {year} {2001}),\ \Eprint
  {https://arxiv.org/abs/math-ph/0103018} {arXiv:math-ph/0103018 [math-ph]}
  \BibitemShut {NoStop}%
\bibitem [{\citenamefont {Kov\'acs}\ \emph
  {et~al.}(2014{\natexlab{c}})\citenamefont {Kov\'acs}, \citenamefont
  {El\ifmmode~\mbox{\c{c}}\else \c{c}\fi{}i}, \citenamefont {Weigel},\ and\
  \citenamefont {Igl\'oi}}]{kovacsPotts}%
  \BibitemOpen
  \bibfield  {author} {\bibinfo {author} {\bibfnamefont {I.~A.}\ \bibnamefont
  {Kov\'acs}}, \bibinfo {author} {\bibfnamefont {E.~M.}\ \bibnamefont
  {El\ifmmode~\mbox{\c{c}}\else \c{c}\fi{}i}}, \bibinfo {author} {\bibfnamefont
  {M.}~\bibnamefont {Weigel}},\ and\ \bibinfo {author} {\bibfnamefont
  {F.}~\bibnamefont {Igl\'oi}},\ }\bibfield  {title} {\bibinfo {title} {Corner
  contribution to cluster numbers in the potts model},\ }\href
  {https://doi.org/10.1103/PhysRevB.89.064421} {\bibfield  {journal} {\bibinfo
  {journal} {Phys. Rev. B}\ }\textbf {\bibinfo {volume} {89}},\ \bibinfo
  {pages} {064421} (\bibinfo {year} {2014}{\natexlab{c}})}\BibitemShut
  {NoStop}%
\bibitem [{\citenamefont {Lorenz}\ and\ \citenamefont
  {Ziff}(1998{\natexlab{a}})}]{perc_cr}%
  \BibitemOpen
  \bibfield  {author} {\bibinfo {author} {\bibfnamefont {C.~D.}\ \bibnamefont
  {Lorenz}}\ and\ \bibinfo {author} {\bibfnamefont {R.~M.}\ \bibnamefont
  {Ziff}},\ }\bibfield  {title} {\bibinfo {title} {Precise determination of the
  bond percolation thresholds and finite-size scaling corrections for the sc,
  fcc, and bcc lattices},\ }\href {https://doi.org/10.1103/PhysRevE.57.230}
  {\bibfield  {journal} {\bibinfo  {journal} {Phys. Rev. E}\ }\textbf {\bibinfo
  {volume} {57}},\ \bibinfo {pages} {230} (\bibinfo {year}
  {1998}{\natexlab{a}})}\BibitemShut {NoStop}%
\bibitem [{\citenamefont {Lorenz}\ and\ \citenamefont
  {Ziff}(1998{\natexlab{b}})}]{perc_cr2}%
  \BibitemOpen
  \bibfield  {author} {\bibinfo {author} {\bibfnamefont {C.~D.}\ \bibnamefont
  {Lorenz}}\ and\ \bibinfo {author} {\bibfnamefont {R.~M.}\ \bibnamefont
  {Ziff}},\ }\bibfield  {title} {\bibinfo {title} {Universality of the excess
  number of clusters and the crossing probability function in three-dimensional
  percolation},\ }\href {https://doi.org/10.1088/0305-4470/31/40/009}
  {\bibfield  {journal} {\bibinfo  {journal} {J. Phys. A}\ }\textbf {\bibinfo
  {volume} {31}},\ \bibinfo {pages} {8147} (\bibinfo {year}
  {1998}{\natexlab{b}})}\BibitemShut {NoStop}%
\bibitem [{\citenamefont {Wang}\ \emph {et~al.}(2013)\citenamefont {Wang},
  \citenamefont {Zhou}, \citenamefont {Zhang}, \citenamefont {Garoni},\ and\
  \citenamefont {Deng}}]{note1}%
  \BibitemOpen
  \bibfield  {author} {\bibinfo {author} {\bibfnamefont {J.}~\bibnamefont
  {Wang}}, \bibinfo {author} {\bibfnamefont {Z.}~\bibnamefont {Zhou}}, \bibinfo
  {author} {\bibfnamefont {W.}~\bibnamefont {Zhang}}, \bibinfo {author}
  {\bibfnamefont {T.~M.}\ \bibnamefont {Garoni}},\ and\ \bibinfo {author}
  {\bibfnamefont {Y.}~\bibnamefont {Deng}},\ }\bibfield  {title} {\bibinfo
  {title} {Bond and site percolation in three dimensions},\ }\href
  {https://doi.org/10.1103/PhysRevE.87.052107} {\bibfield  {journal} {\bibinfo
  {journal} {Phys. Rev. E}\ }\textbf {\bibinfo {volume} {87}},\ \bibinfo
  {pages} {052107} (\bibinfo {year} {2013})}\BibitemShut {NoStop}%
\bibitem [{SI()}]{SI}%
  \BibitemOpen
  \href@noop {} {}\bibinfo {note} {See Supplemental Material at [LINK] for
  Figures S1-S4.}\BibitemShut {Stop}%
\bibitem [{\citenamefont {Kasteleyn}\ and\ \citenamefont
  {Fortuin}(1969)}]{Fortuin-Kasteleyn}%
  \BibitemOpen
  \bibfield  {author} {\bibinfo {author} {\bibfnamefont {P.~W.}\ \bibnamefont
  {Kasteleyn}}\ and\ \bibinfo {author} {\bibfnamefont {C.~M.}\ \bibnamefont
  {Fortuin}},\ }\bibfield  {title} {\bibinfo {title} {Phase transitions in
  lattice systems with random local properties},\ }\href@noop {} {\bibfield
  {journal} {\bibinfo  {journal} {J. Phys. Soc. Japan}\ }\textbf {\bibinfo
  {volume} {26}},\ \bibinfo {pages} {11} (\bibinfo {year} {1969})}\BibitemShut
  {NoStop}%
\bibitem [{\citenamefont {Kov\'acs}\ and\ \citenamefont
  {Igl\'oi}(2012{\natexlab{a}})}]{Kovacs2012gap}%
  \BibitemOpen
  \bibfield  {author} {\bibinfo {author} {\bibfnamefont {I.~A.}\ \bibnamefont
  {Kov\'acs}}\ and\ \bibinfo {author} {\bibfnamefont {F.}~\bibnamefont
  {Igl\'oi}},\ }\bibfield  {title} {\bibinfo {title} {Universal logarithmic
  terms in the entanglement entropy of 2d, 3d and 4d random transverse-field
  ising models.},\ }\href {https://doi.org/10.1209/0295-5075/97/67009}
  {\bibfield  {journal} {\bibinfo  {journal} {EPL}\ }\textbf {\bibinfo {volume}
  {97}},\ \bibinfo {pages} {67009} (\bibinfo {year}
  {2012}{\natexlab{a}})}\BibitemShut {NoStop}%
\bibitem [{\citenamefont {Wu}(1982)}]{wu}%
  \BibitemOpen
  \bibfield  {author} {\bibinfo {author} {\bibfnamefont {F.~Y.}\ \bibnamefont
  {Wu}},\ }\bibfield  {title} {\bibinfo {title} {The {P}otts model},\ }\href
  {https://doi.org/10.1103/RevModPhys.54.235} {\bibfield  {journal} {\bibinfo
  {journal} {Rev. Mod. Phys.}\ }\textbf {\bibinfo {volume} {54}},\ \bibinfo
  {pages} {235} (\bibinfo {year} {1982})}\BibitemShut {NoStop}%
\bibitem [{\citenamefont {Cardy}(2005)}]{CG}%
  \BibitemOpen
  \bibfield  {author} {\bibinfo {author} {\bibfnamefont {J.}~\bibnamefont
  {Cardy}},\ }\bibfield  {title} {\bibinfo {title} {{SLE} for theoretical
  physicists},\ }\href {https://doi.org/doi.org/10.1016/j.aop.2005.04.001}
  {\bibfield  {journal} {\bibinfo  {journal} {Ann. Phys.}\ }\textbf {\bibinfo
  {volume} {318}},\ \bibinfo {pages} {81} (\bibinfo {year} {2005})},\ \bibinfo
  {note} {special Issue}\BibitemShut {NoStop}%
\bibitem [{\citenamefont {Kov\'acs}\ and\ \citenamefont
  {Igl\'oi}(2012{\natexlab{b}})}]{EPL}%
  \BibitemOpen
  \bibfield  {author} {\bibinfo {author} {\bibfnamefont {I.~A.}\ \bibnamefont
  {Kov\'acs}}\ and\ \bibinfo {author} {\bibfnamefont {F.}~\bibnamefont
  {Igl\'oi}},\ }\bibfield  {title} {\bibinfo {title} {Universal logarithmic
  terms in the entanglement entropy of 2d, 3d and 4d random transverse-field
  ising models},\ }\href {https://doi.org/10.1209/0295-5075/97/67009}
  {\bibfield  {journal} {\bibinfo  {journal} {EPL}\ }\textbf {\bibinfo {volume}
  {97}},\ \bibinfo {pages} {67009} (\bibinfo {year}
  {2012}{\natexlab{b}})}\BibitemShut {NoStop}%
\end{thebibliography}%

\end{document}


\title{Supplemental Material:\\Cluster tomography in percolation}
\author{Helen S. Ansell}
\affiliation{Department of Physics and Astronomy, Northwestern University, Evanston, IL 60208}
\author{Samuel J. Frank}
\affiliation{Department of Physics and Astronomy, Northwestern University, Evanston, IL 60208}
\author{Istv\'an A. Kov\'acs}
\affiliation{Department of Physics and Astronomy, Northwestern University, Evanston, IL 60208}
\affiliation{Northwestern Institute on Complex Systems, Northwestern University, Evanston, IL 60208}
\affiliation{Department of Engineering Sciences and Applied Mathematics, Northwestern University, Evanston, IL 60208}

\date{\today}

\maketitle

\renewcommand{\thefigure}{\arabic{figure}}
\renewcommand{\theHfigure}{ \arabic{figure}}
\makeatletter
\renewcommand{\fnum@figure}{FIG. S\thefigure}
\makeatother

\newpage
\section*{Supplemental Figures}

\begin{figure}[h]
\includegraphics[width=0.8\textwidth]{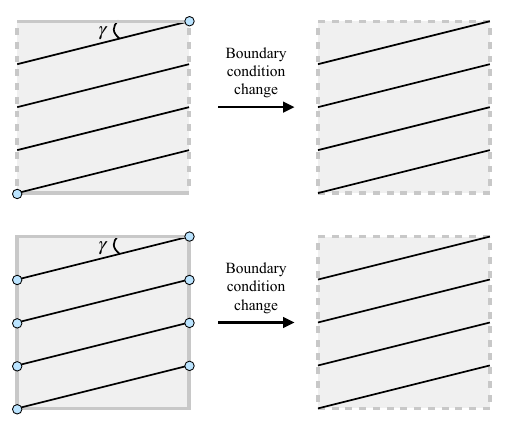}
\caption{(Top) For line type 6, where the endpoints are on opposite edges of the square, we can impose PBC on the other two edges to numerically determine the endpoint contribution to high precision. Changing the FBC on the top and bottom to PBC gives a closed loop, with no endpoints, along the same set of sites. (Bottom) For line type 8 the initial configuration (left) requires FBC on all sides. For $\gamma = \arctan(1/n)$, a closed loop with no endpoints can be made after imposing PBC by placing two type 8 lines and $n-2$ type 6 lines in the FBC system. The corner contribution to the cluster number count can be determined by counting the contribution along each of the lines in the FBC system, subtracting the total count along the closed loop after applying PBC, and then subtracting off the known contribution due to the $n-2$ type 6 lines. Here solid grey lines represent free system boundaries while dashed lines represent PBC. Blue circles indicate the presence of line endpoints.}
\end{figure}

\begin{figure}[h]
\includegraphics[width=\textwidth]{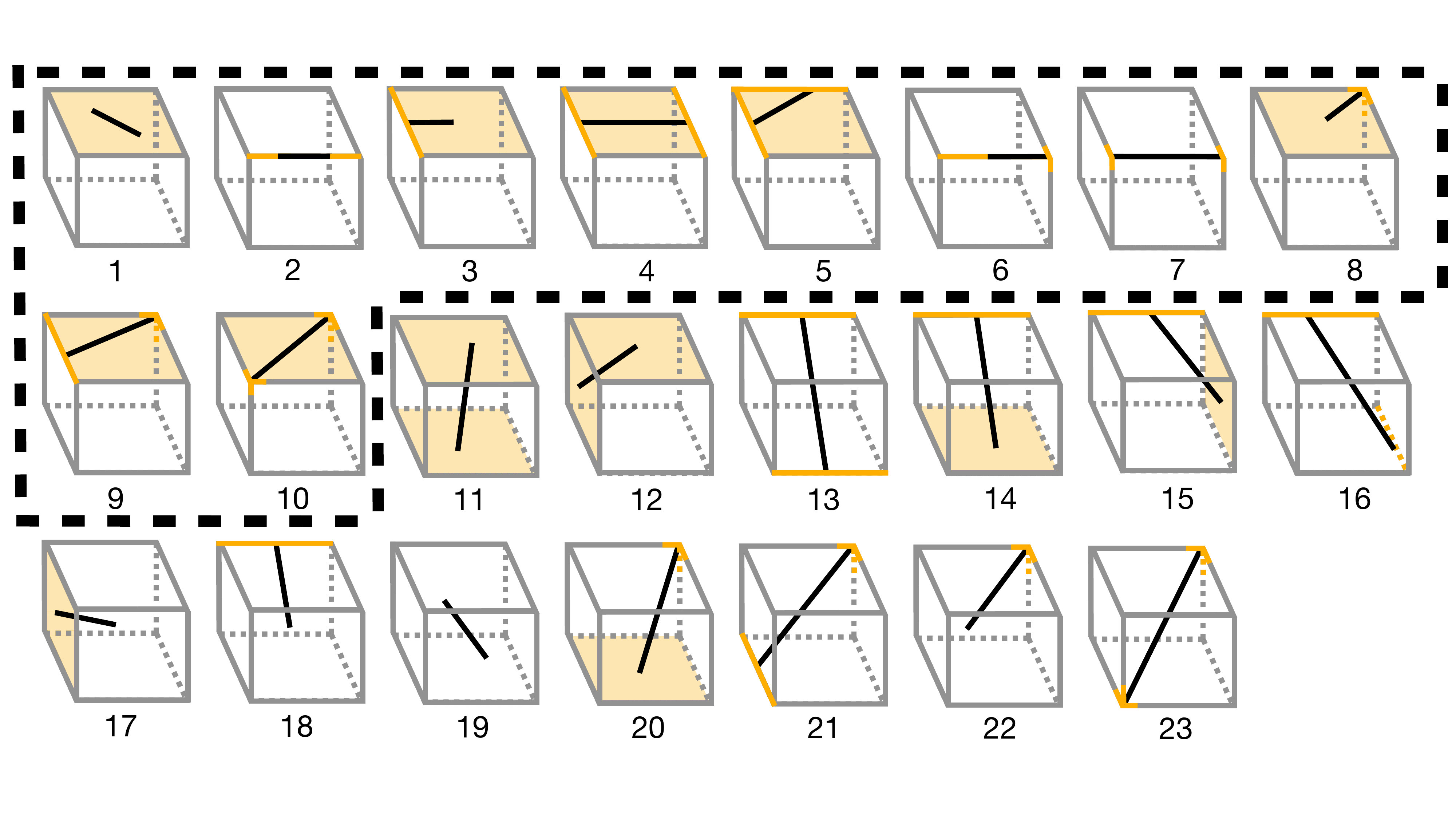}
\caption{Summary of the 23 possible line configurations in a 3$d$ cubic system with endpoints touching faces, edges, corners, and the bulk of the cube considered separately. The first ten configurations are equivalent to the line types considered in 2$d$.}
\end{figure}

\begin{figure}[h]
\centering
\includegraphics[width=\textwidth]{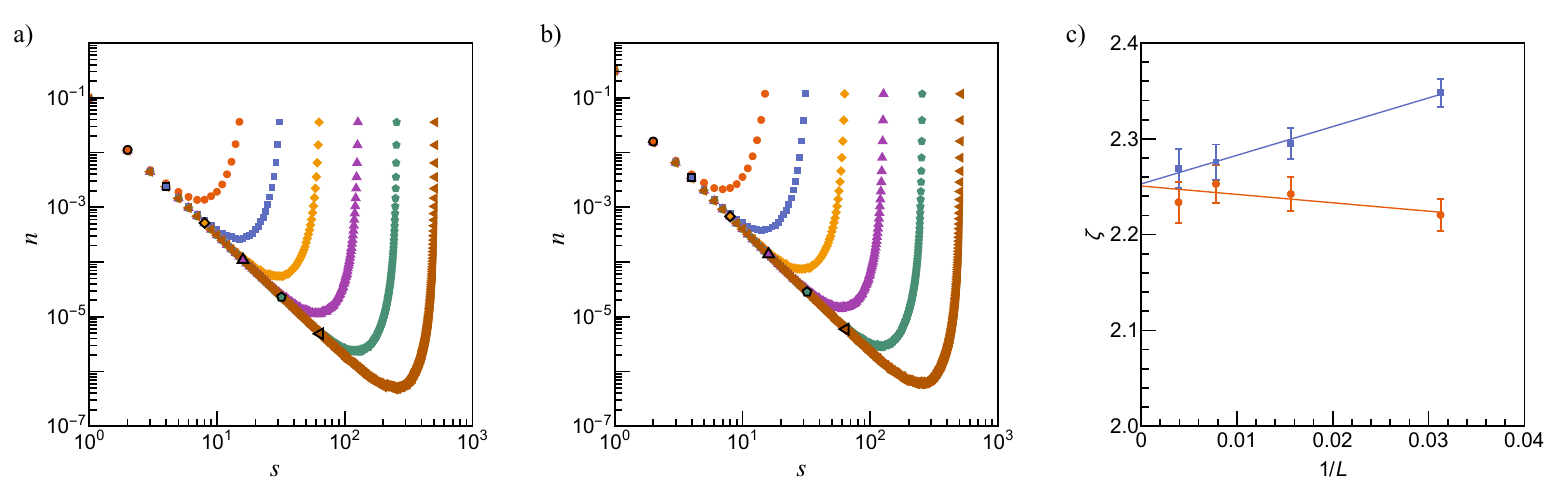}
\caption{(a-b) Gap size statistics $n(s)\sim s^{-\zeta}$ for clusters along lines of length $L$ on the face of a cube with periodic boundary conditions along two axes and free boundary conditions in the third for (a) site percolation and (b) bond percolation. (c) Size-dependence of the $\zeta$ exponent, calculated from two-point fits using the highlighted points in a and b. The extrapolated $\zeta$ value is $2.25(1)$.}
\label{fig:gaps}
\end{figure}

\begin{figure}[h]
\centering
\includegraphics[width=\textwidth]{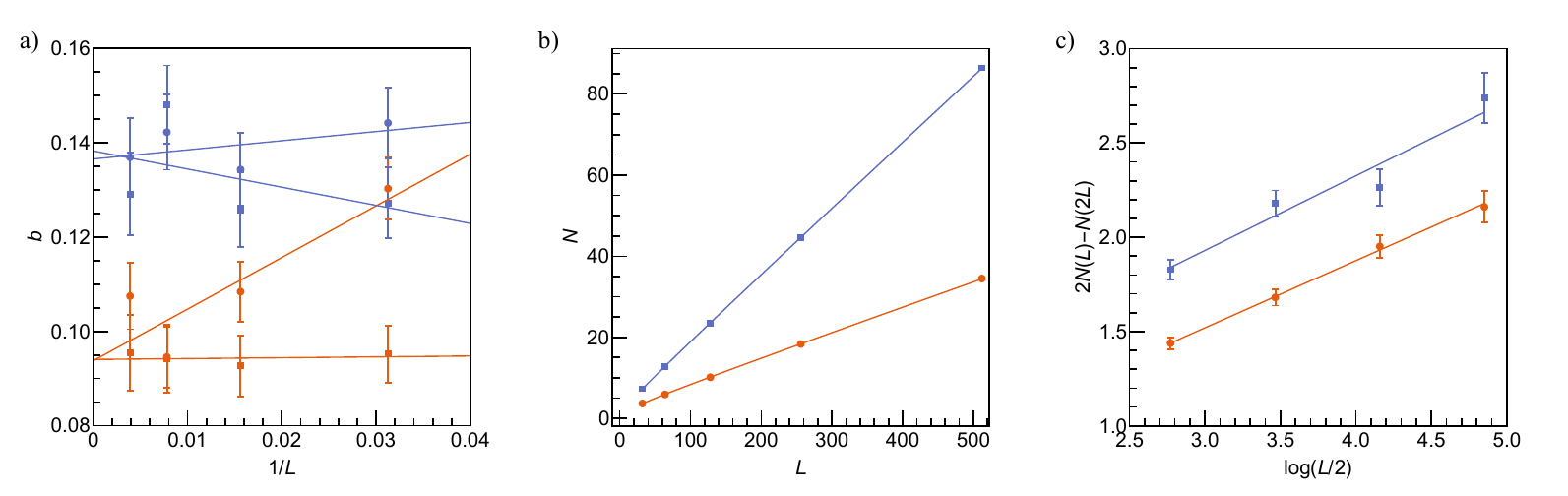}
\caption{Numerical determination of the cluster count exponent $b$ without relying on the high precision numerical techniques. 
(a) Finite-size scaling of $b_{bb}$ (red) and $b_{ss}$ (blue) calculated in a square system with free boundaries all around without using the boundary changing technique. Points are shown for site percolation (circles) and bond percolation (squares) with 10 000 samples at each $L$. 
The extrapolated $b$ values are $b_{bb} = 0.094(6)$ and $b_{ss} = 0.137(7)$.
(b) Number of clusters $N$ along a line with two traversing endpoints with $\gamma = \pi/2$ spanning a system of length $L$ with free boundaries on all sides for site percolation (red circles) and bond percolation (blue squares).
(c) Estimating the corner contribution of the points in b by comparing points at successive sizes to cancel the area law on average. Averaging the slopes of the fit lines for site percolation (red circles) and bond percolation (blue squares) gives $b_{tt} = 0.38(6)$, consistent with the result obtained using high precision techniques.
}
\label{fig:free-boundaries}
\end{figure}